\documentclass[a4paper,fleqn]{cas-sc}

\usepackage[numbers]{natbib}

\usepackage{algorithmic}
\usepackage{algorithm}

\usepackage{subfigure}
\usepackage{graphicx}
\usepackage{bbding}

\def\tsc#1{\csdef{#1}{\textsc{\lowercase{#1}}\xspace}}
\tsc{WGM}
\tsc{QE}
\begin{document}
\let\WriteBookmarks\relax
\def\floatpagepagefraction{1}
\def\textpagefraction{.001}

\shorttitle{IoT-AMLHP}    

% \shortauthors{<short author list for running head>}  

\title [mode = title]{IoT-AMLHP: Aligned Multimodal Learning of Header-Payload Representations for Resource-Efficient Malicious IoT Traffic Classification}  

% \tnotemark[<tnote number>] 

% \tnotetext[]{This work was supported by the National Key R\&D Program of China (Grants No. 2021QY0700), the National Natural Science Foundation of China (Grants No. U1836104, 61602247, 61801073, 61931004, 62072250).} 


% \author[<aff no>]{<author name>}[<options>]

% \cormark[<corr mark no>]

% \fnmark[<footnote mark no>]

% \ead{lwwnjust@njust.edu.cn}

% Author
\author[1]{Fengyuan Nie}
\author[2,3]{Guangjie Liu}
\cormark[1] 
\cortext[1]{Corresponding author: Guangjie Liu} 
% \cortext[]{E-mail address: niefengyuan@njust.edu.cn (F. Nie)}
\cortext[0]{niefengyuan@njust.edu.cn (F. Nie); gjieliu@gmail.com (G. Liu); lwwnjust@njust.edu.cn (W. Liu); jiananwong@njust.edu.cn (J. Huang); njustgb565@163.com (B. Gao)}
\author[1]{Weiwei Liu}
% \cormark[1] 
\author[1]{Jianan Huang}
\author[1]{Bo Gao}

\address[1]{School of Automation, Nanjing University of Science and Technology, Nanjing 210094, China}
\address[2]{School of Electronics and information engineering, Nanjing University of Information Science and Technology, Nanjing 210044, China}
\address[3]{Key Laboratory of Intelligent Support Technology for Complex Environments, Ministry of Education, Nanjing University of Information Science and Technology, Nanjing, 210044, Jiangsu, China}

% \ead[url]{<URL>}

% \credit{<Credit authorship details>}

% \affiliation[<aff no>]{organization={},
% %          citysep={}, % Uncomment if no comma needed between city and postcode

% \author[<aff no>]{<author name>}[<options>]

% Footnote of the second author

% % Email id of the second author
% \ead{}

% % URL of the second author

% % Credit authorship

% % Address/affiliation
% \affiliation[<aff no>]{organization={},
% %          citysep={}, % Uncomment if no comma needed between city and postcode

% % Corresponding author text
% \cortext[1]{Corresponding author}

% % Footnote text


\begin{abstract}
Traffic classification is crucial for securing Internet of Things (IoT) networks. Deep learning-based methods can autonomously extract latent patterns from massive network traffic, demonstrating significant potential for IoT traffic classification tasks. However, the limited computational and spatial resources of IoT devices pose challenges for deploying more complex deep learning models. Existing methods rely heavily on either flow-level features or raw packet byte features. Flow-level features often require inspecting entire or most of the traffic flow, leading to excessive resource consumption, while raw packet byte features fail to distinguish between headers and payloads, overlooking semantic differences and introducing noise from feature misalignment. Therefore, this paper proposes IoT-AMLHP, an aligned multimodal learning framework for resource-efficient malicious IoT traffic classification. Firstly, the framework constructs a packet-wise header-payload representation by parsing packet headers and payload bytes, resulting in an aligned and standardized multimodal traffic representation that enhances the characterization of heterogeneous IoT traffic. Subsequently, the traffic representation is fed into a resource-efficient neural network comprising a multimodal feature extraction module and a multimodal fusion module. The extraction module employs efficient depthwise separable convolutions to capture multi-scale features from different modalities while maintaining a lightweight architecture. The fusion module adaptively captures complementary features from different modalities and effectively fuses multimodal features. Extensive experiments on three public IoT traffic datasets demonstrate that the proposed IoT-AMLHP outperforms state-of-the-art methods in classification accuracy while significantly reducing computational and spatial resource overhead, making it highly suitable for deployment in resource-constrained IoT environments.

\end{abstract}


% \begin{highlights}
% \item 
% \item 
% \item 
% \end{highlights}

% Keywords

\maketitle

% Main text
\section{Introduction}\label{}
The development of the Internet of Things (IoT) is progressing at an unprecedented pace, with an ever-increasing number of smart devices connecting to the Internet \cite{IoT_1,IoT_2,IoT_3}. By 2025, the global number of IoT devices is expected to exceed 75 billion \cite{750}. This explosive growth in device count has led to a massive surge in network traffic, which encompasses not only legitimate application traffic and cloud service traffic but also a steadily rising volume of malicious network traffic, such as DDoS attacks and botnet activities \cite{DDos,Botnet}. The malicious traffic can not only result in user privacy breaches but also cause significant financial losses. Therefore, network traffic classification techniques are crucial for maintaining the security of the IoT networks.

Early network traffic classification methods primarily relied on port-based and payload-based approaches \cite{p2_1}. However, the widespread adoption of dynamic port allocation and network address translation (NAT) technologies gradually diminished the effectiveness of port-based methods \cite{p2_2}. Payload-based methods, such as deep packet inspection (DPI), which match patterns and keywords within the traffic, may be more reliable than port-based approaches \cite{p2_3_1,p2_3_2}. Nonetheless, the growing prevalence of encrypted traffic in IoT networks, coupled with attackers' ability to obscure easily detectable fields by emulating normal traffic behavior, has resulted in a decline in the performance of payload-based methods \cite{c2,c3}. Machine learning-based traffic classification methods, such as random forests and support vector machines, have achieved promising results in identifying IoT traffic, as they do not rely on specific ports or keywords \cite{c1}. However, these methods often place excessive emphasis on feature engineering, making it challenging to maintain classification accuracy as network traffic evolves \cite{cmftc}. Moreover, the limited learning abilities of machine learning models impose performance constraints on classification accuracy.

Recently, deep learning-based methods have received significant attention due to their powerful automatic feature extraction capability, providing a new reliable solution for maintaining network security \cite{11}. Compared with traditional machine learning-based methods, deep learning-based methods can automatically mine high-dimensional feature representations from massive network traffic data, capturing the inherent deep patterns within the data, thereby significantly improving the ability to identify malicious traffic \cite{tifs}. However, deep learning-based methods typically require substantial computational and spatial resources for model training, inference, and deployment, posing significant challenges for deployment in resource-constrained IoT environments. Consequently, researching lightweight and resource-efficient traffic identification methods has become a current research hotspot.

Additionally, most machine learning and deep learning-based traffic identification methods are typically classified into flow-level and packet-level methods, depending on the input traffic sample types. Flow-level methods generally utilize flow-level features (such as flow duration and average packet length) and raw flow bytes as model inputs  \cite{appscanner,CUMUL}. However, they require inspecting all or most session data, increasing computational and storage resource consumption and introducing higher detection latency \cite{SAM}. In contrast, packet-level methods offer better real-time performance, but they typically directly use raw packet bytes as inputs, ignoring the semantic and pattern inconsistencies between packet headers and payloads, as well as the feature misalignment caused by the diversity of protocol formats and packet lengths.

In this paper, we propose IoT-AMLHP, an aligned multimodal learning framework for resource-efficient malicious IoT traffic classification. IoT-AMLHP is capable of processing more fine-grained multimodal traffic features at the packet level, rather than flow-level features, and outputting classification results as packets arrive, thereby meeting the low-latency requirements for traffic classification. Specifically, IoT-AMLHP first constructs a packet-wise representation by parsing the headers and payloads of incoming traffic packets, resulting in an aligned and standardized multimodal traffic feature. Subsequently, a resource-efficient neural network is designed to extract and fuse these multimodal traffic features. The neural network comprises a multimodal feature extraction module and a multimodal fusion module. The multimodal feature extraction module employs depthwise separable convolutions to further process the representations, effectively mining multi-scale protocol field features and payload semantic features from the packets while maintaining a lightweight structure with lower computational complexity. Next, the multimodal fusion module utilizes a multi-head self-attention (MHSA) mechanism to dynamically focus on complementary features across different modalities and achieve optimal fusion, thereby enhancing the model's identification capability. Finally, a softmax function in the classifier makes decisions on each packet, achieving accurate identification of malicious IoT traffic.

The main contributions of this study can be summarized as follows:
\begin{itemize}
\item{We propose an aligned multimodal learning framework, IoT-AMLHP, for resource-efficient malicious IoT traffic classification. The framework constructs an aligned and standardized packet header-payload representation by parsing traffic packets, which is then fed into a resource-efficient neural network. IoT-AMLHP achieves high accuracy and low latency in classifying malicious IoT traffic while effectively reducing the consumption of computational and spatial resources, making it suitable for deployment in resource-constrained IoT environments.}
\item{We construct a packet-wise header-payload representation that employs an aligned and compact representation of packet header and payload features. This not only effectively represents the structural information and semantic content within headers and payloads but also mitigates the issue of feature misalignment, thereby enhancing model performance and generalizability.}
\item{We design a resource-efficient neural network for efficient extraction and fusion of multimodal traffic features. This neural network comprises a multimodal feature extraction module and a multimodal fusion module. The multimodal feature extraction module uses a depthwise separable convolutional network as the backbone to reduce the model complexity and parameters. The multimodal fusion module leverages a multi-head self-attention mechanism to effectively fuse features from different modalities, further improving the accuracy of malicious traffic classification.}
\item{We evaluate the proposed IoT-AMLHP on three typical public IoT traffic datasets and compare it with various state-of-the-art baseline methods. Extensive experimental results demonstrate the superiority of our method in terms of accuracy and resource efficiency in malicious IoT traffic classification.}
\end{itemize}

The structure of the remainder of this paper is outlined as follows: Section \ref{sec::related_work} provides an overview of the related work. The architecture of the proposed IoT-AMLHP is detailed in Section \ref{sec::methodology}. Experimental setups, along with a comprehensive discussion of the results, are presented in Section \ref{sec::experiments}. Finally, the conclusions are drawn in Section \ref{sec::conclusions}.

\section{Related work} \label{sec::related_work}
With the advent of the IoT era, the number of IoT devices has experienced exponential growth, accompanied by various network communication activities, resulting in massive network traffic generation. Hidden within this massive network traffic data are potential malicious activities, posing significant threats to the security of IoT networks.

Early network traffic classification methods were primarily categorized into port-based and payload-based methods. Port-based methods identify traffic types by inspecting the source and destination ports of packets. However, with the abuse of ports and the widespread adoption of NAT technologies, the efficacy of port-based techniques has considerably diminished. To overcome these limitations, researchers proposed payload-based DPI techniques. DPI is a technology that identifies traffic types by examining the content of packets. Although DPI techniques have achieved successful applications in some industrial products \cite{DPI_1,DPI_2}, recent studies have revealed that, due to the widespread use of encryption in IoT communications, DPI methods also face significant challenges \cite{DPI_c1,DPI_c2}.

Machine learning methods have received widespread attention from researchers due to their excellent pattern recognition capabilities. In \cite{maniriho2020anomaly}, Maniriho et al. introduced a detection method utilizing the random forest (RF) algorithm. This method employs a hybrid feature selection engine to identify the most relevant features, which are subsequently used to train a random forest model for traffic classification within IoT networks, achieving high identification accuracy for abnormal traffic on the IoTID20 dataset. Shafiq et al. \cite{shafiq2020corrauc} proposed a novel framework employing a CorrAUC feature selection algorithm for accurate malicious traffic detection in IoT networks, with their method achieving over 96\% performance on the BoT-IoT dataset. Shi et al. \cite{shi2021three} introduced a three-layer hybrid classification model designed for smart home environments, which uses a two-layer feature processing approach, integrating principal component analysis (PCA) and random forest to minimize information loss. It classifies four common attacks using four binary classifiers that identify various traffic behaviors. However, these machine learning-based methods typically require complex feature engineering to improve their ability to detect malicious traffic. Furthermore, due to the shallow learning capabilities of these machine learning models, their detection performance is limited when handling large and heterogeneous IoT traffic data.

Compared to traditional machine learning-based methods, deep learning-based methods can autonomously learn distinctive feature representations from massive amounts of network traffic, not only reducing the complexity of manual feature engineering but also achieving superior detection performance. In \cite{adaptiver2024deep}, an adaptive deep learning model that integrates convolutional neural networks (CNN) with gated recurrent units (GRU) is proposed to analyze network traffic patterns and detect botnet activities. Yao et al. \cite{ATTLSTM} proposed a network traffic classification method that models time-series traffic using recurrent neural networks (RNN) and incorporates attention mechanisms through long short term memory (LSTM) and hierarchical attention networks (HAN), achieving improved accuracy compared to traditional methods. In \cite{tscrnn}, a spatiotemporal-based traffic classification method, TSCRNN, is proposed, which combines CNN for spatial feature extraction and LSTM for temporal pattern learning to identify traffic in IIoT networks. Zhu et al. \cite{cmtsnn} proposed a multi-classification deep learning model for detecting IoT traffic, which incorporates a cost penalty matrix for dataset preprocessing, extracts time-series and spatial features for robust representation, and applies an improved loss function to address data imbalance and enhance multi-classification accuracy.

Although deep learning-based methods have demonstrated outstanding performance in identifying malicious IoT traffic, they typically require substantial computational and spatial resources, posing challenges for their deployment in resource-constrained IoT environments. Consequently, some researchers have proposed resource-efficient methods to enhance the model's lightweight performance and reduce computational complexity. Zhao et al. \cite{zhao2021novel} introduced a lightweight neural network approach for IoT intrusion detection, employing PCA for traffic feature dimensionality reduction. Their network architecture with expansion, compression structure, and channel shuffle operations enabled effective yet low-complexity feature extraction. A NID loss function enhanced multi-class performance. Their method achieved high accuracy with low computational cost on the two traffic datasets. In \cite{wang2024spatial}, a knowledge distillation framework for network anomaly detection is presented. They developed a multi-scale spatial-temporal residual network as the teacher model to extract features from network traffic, and then distilled the knowledge to a lightweight student model for malicious traffic detection. In \cite{matec}, a lightweight traffic classification model is proposed to efficiently handle network traffic by utilizing an improved Transformer network, significantly reducing the number of parameters while improving classification accuracy and efficiency compared to other methods.

\begin{table}[htbp]
\label{Tab:1}
\renewcommand{\arraystretch}{1.0}
\centering
\caption{A summary and comparison of existing traffic classification methods.}
\resizebox{\linewidth}{!}{
\begin{tabular}{cc>{\centering\arraybackslash}m{3cm}>{\centering\arraybackslash}m{3.3cm}>{\centering\arraybackslash}m{3.5cm}c>{\centering\arraybackslash}m{2cm}>{\centering\arraybackslash}m{2cm}>{\centering\arraybackslash}m{2cm}}
\toprule
\textbf{Ref ID.}   & \textbf{Type} & \textbf{Hierarchical Level of the Utilized Feature}  & \textbf{Representation Method}                     & \textbf{Learning Architecture} & \textbf{Input Size} & \textbf{Feature Alignment}  & \textbf{Multimodal Learning} & \textbf{Lightweight} \\ \midrule[1pt]
\cite{maniriho2020anomaly}  & ML   & Flow-Level   & Flow Statistical Features and Flags & Random Forest (RF) & $-$          & $-$   & \XSolidBrush                 & \XSolidBrush                  \\
\cite{shafiq2020corrauc}  & ML   & Flow-Level   & Flow Statistical Features           & Decision Tree (DT) & $-$          & $-$       & \XSolidBrush             & \XSolidBrush                  \\
\cite{shi2021three}  & ML   & Flow-Level   & Flow Statistical Features and Flags & PCA$+$RF & $-$          & $-$       & \XSolidBrush            & \XSolidBrush                  \\
\cite{adaptiver2024deep}  & DL   & Flow-Level   & Flow Statistical Features and Flags          & CNN$+$GRU & $-$          & $-$        & \XSolidBrush            & \XSolidBrush                  \\
\cite{ATTLSTM}  & DL   & Flow-Level   & Raw Flow Bytes                     & LSTM$+$HAN & 10$\times$1500    & \XSolidBrush       & \XSolidBrush            & \XSolidBrush                  \\
\cite{tscrnn}  & DL   & Flow-Level   & Raw Flow Bytes                     & CNN$+$LSTM & 15$\times$1500    & \XSolidBrush        & \XSolidBrush           & \XSolidBrush                  \\
\cite{cmtsnn}  & DL   & Flow-Level   & Raw Flow Bytes                     & CNN$+$BILSTM & 10$\times$1000    & \XSolidBrush       & \XSolidBrush            & \XSolidBrush                  \\
\cite{zhao2021novel}  & DL   & Flow-Level   & Flow Statistical Features and Flags            & PCA$+$CNN & $-$          & $-$        & \XSolidBrush           & \Checkmark                  \\
\cite{wang2024spatial}  & DL   & Flow-Level   & Flow Statistical Features and Flags         & 1DCNN$+$LSTM & $-$          & $-$      & \XSolidBrush             & \Checkmark                  \\
\cite{matec}  & DL   & Flow-Level   & Raw Flow Bytes                     & Transformer & 3$\times$786      & \XSolidBrush         & \XSolidBrush           & \Checkmark                  \\
\cite{Deeppacket}  & DL   & Packet-Level & Raw Packet Bytes                   & 1DCNN, SAE & 1500       & \XSolidBrush    & \XSolidBrush               & \XSolidBrush                  \\
\cite{LSTM}  & DL   & Packet-Level & Raw Packet Bytes                   & 1DCNN, 2DCNN, LSTM & 784        & \XSolidBrush     & \XSolidBrush               & \XSolidBrush                  \\ \midrule[1pt]
IoT-AMLHP & DL   & Packet-Level & Header-Payload Representation      & Depthwise Separable 1DCNN$+$MHSA & 128, 64     & \Checkmark      & \Checkmark              & \Checkmark                  \\ \bottomrule
\end{tabular}
}
\end{table}

Despite the high accuracy achieved by these machine learning and deep learning-based approaches in traffic classification tasks, they predominantly rely on flow-level statistical features or raw flow byte sequences as model inputs, as summarized in Table 1. Extracting these flow-level features usually requires inspecting all or most of the traffic flow to be computed. Accurate features or a sufficient number of bytes can only be extracted after the traffic flow has been active for some time or even completed, not only causing additional computational and storage overhead but also lacking real-time performance. Consequently, packet-level traffic identification methods have been proposed \cite{Deeppacket} \cite{LSTM}. However, these methods still directly use the raw byte features of packets as model inputs, which not only ignores the semantic and pattern differences between packet header fields and payload bytes but also overlooks the noise introduced by feature misalignment caused by significant differences in packet lengths and formats across different protocols. 

To overcome these challenges, we propose IoT-AMLHP, a resource-efficient malicious traffic classification framework for IoT networks. As summarized in Table 1, IoT-AMLHP differs from prior works across dimensions like representation method, input size, feature alignment, multimodal learning, and lightweight design. It operates at the packet level using a compact header-payload representation. By separately parsing headers and payloads, it constructs an aligned multimodal representation that better captures their distinct semantics and mitigates feature misalignment. Moreover, IoT-AMLHP applies multimodal learning to effectively extract and fuse complementary features from both modalities. To ensure efficiency without sacrificing accuracy, it integrates lightweight architectural components such as depthwise separable convolutions, making it well-suited for deployment in resource-constrained IoT environments.

\begin{figure}[htbp]
    \centering
    \includegraphics[width=0.6\linewidth]{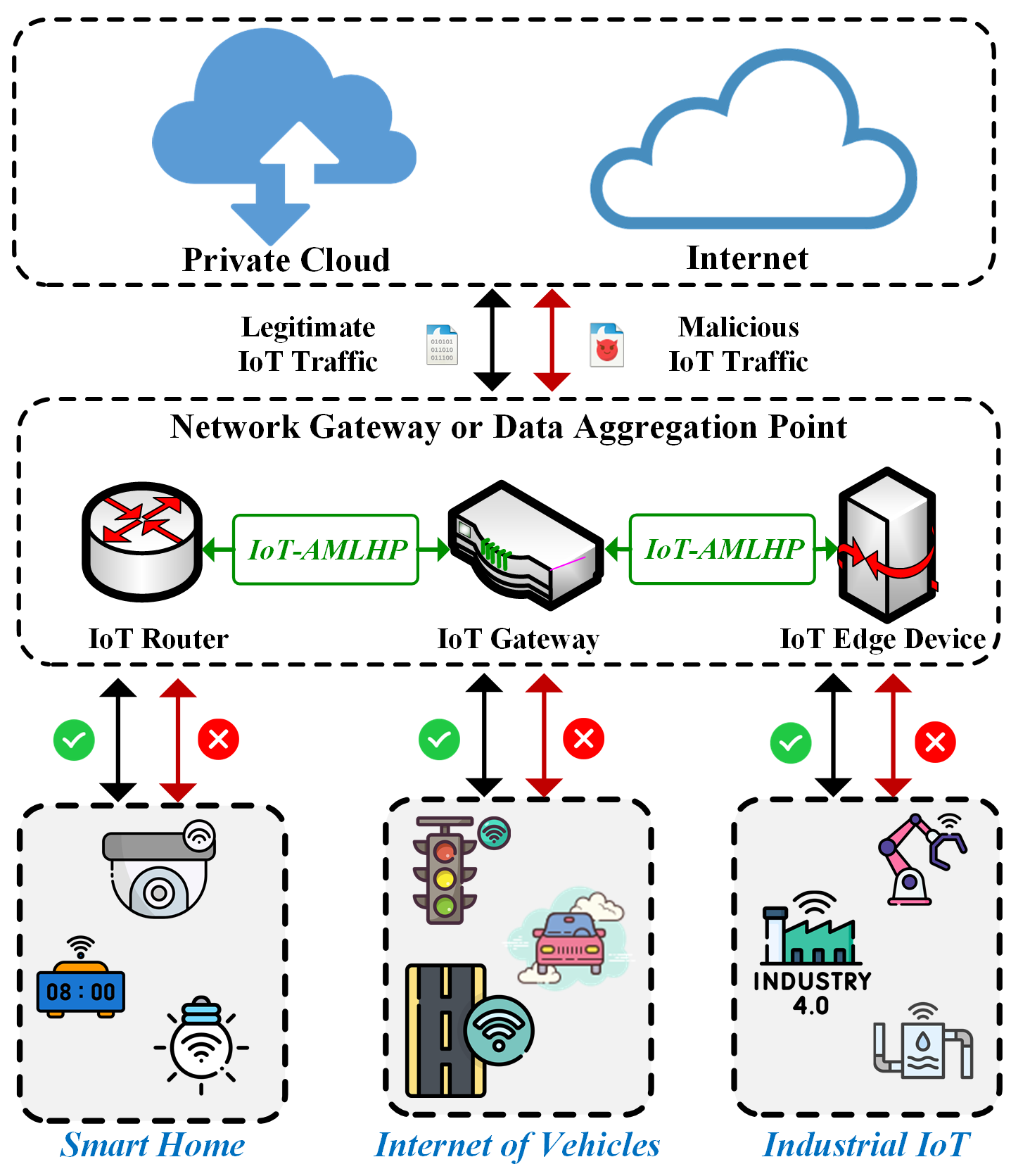}
    % \caption{\color{blue}The illustration of the implementation and deployment of the proposed IoT anomaly detection method.\color{black}}
    \caption{The illustration of the deployment framework of the proposed IoT-AMLHP.}
    \label{IoT_framework}
\end{figure}

\begin{figure*}[htbp]
    \centering
    \includegraphics[width=\linewidth]{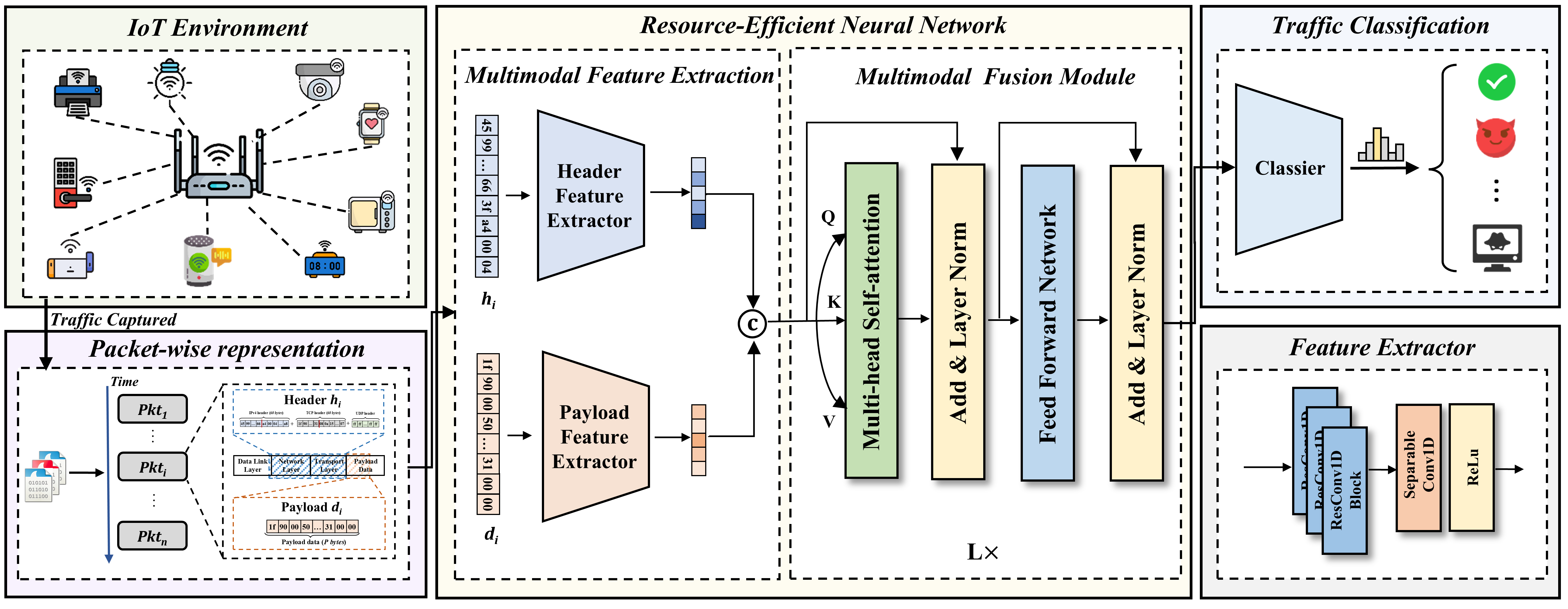}
    \caption{Overview of the proposed IoT-AMLHP.}
    \label{model}
\end{figure*}

\section{Methodology} \label{sec::methodology}
\subsection{Overview}
The proposed IoT-AMLHP is designed as a resource-efficient traffic identification model that can be flexibly deployed on IoT gateways, routers, or edge devices, as illustrated in Fig. 1. It supports monitoring and classifying network traffic generated in IoT environments, such as smart homes, Industrial IoT, and the Internet of Vehicles, during communication with private clouds or the Internet. The framework supports diverse deployment configurations, such as bypass-mode operation on edge devices and direct implementation on IoT gateways, enabling localized processing and real-time traffic analysis. Additionally, its lightweight architecture ensures scalability and low computational overhead, enhancing the security and reliability of IoT networks.

The main workflow of the IoT-AMLHP framework is illustrated in Fig. \ref{model}. First, the captured traffic packets are parsed and processed separately based on the differences in the semantics of the packet header and payload, constructing an aligned and standardized multimodal packet-wise representation. Subsequently, the representation is fed into a resource-efficient neural network comprising a multimodal feature extraction module and a multimodal fusion module. The multimodal feature extraction module employs depth-separable convolutions to mine richer and more discriminative features in multimodal representations from multiple scales, while maintaining a lightweight architecture and lower computational complexity. Next, a multimodal fusion module based on the multi-head self-attention mechanism effectively fuses the extracted head and payload features to generate a comprehensive representation that integrates complementary information from different modalities. The multi-head self-attention mechanism adaptively captures the most discriminative features from different modalities, thus achieving the optimal fusion of multimodal features. Finally, accurate classification of malicious traffic in the IoT environment is achieved through a classifier composed of fully connected layers.

\subsection{Packet-wise header-payload representation}

IoT traffic packets can typically be captured on IoT nodes equipped with sniffing tools as traffic collectors, which are then used to construct an aligned multimodal packet-wise header-payload representation as the input to the model. The representation aims to effectively encode the protocol structures and semantic features of heterogeneous IoT traffic while maintaining a compact and resource-efficient design. In this representation, each incoming packet is first parsed into a hexadecimal byte sequence $pkt_i = [b_1, b_2, ..., b_n]$, where $b_j \in [0x00, 0x\text{ff}]$ and $n$ is the total number of bytes in the packet. Subsequently, the header and payload bytes from each packet $pkt_i$ are separated and parsed, undergoing alignment and standardization operations to better reflect their distinct semantic characteristics.

The packet header representation focuses on capturing protocol-level information such as protocol fields and structures in IoT traffic. These header fields provide insights into the syntactic structure of network protocols and identity information about communicating entities, such as source and destination addresses, ports, and flags, which are widely used in network traffic analysis \cite{nie2024m2vt}. However, directly using raw packet bytes as input can lead to feature misalignment due to varying protocol formats, thereby introducing unnecessary noise that could limit the model's performance \cite{nprint}. To consistently represent header structures across different protocols, we adopt a unified approach that standardizes and aligns header features, as illustrated in Fig. 3. Specifically, the packet header includes data from the network and transport layers while excluding the data link layer, as its primary role is to provide device connection setup and MAC address information, which are less relevant to traffic classification \cite{Datanet}.

After excluding data link layer information, the packet header primarily consists of the network and transport layer protocol headers. By aligning and standardizing the byte data of these headers, we obtain a fixed-length representation for the header. Specifically, at the network layer, we process IPv4 packets, which dominate most network environments \cite{IPv4}. The IPv4 header comprises a 20-byte basic protocol field and a 40-byte optional field, totaling 60 bytes. At the transport layer, we focus on TCP and UDP, which are the primary protocols. The TCP header consists of a 20-byte basic protocol field and a 40-byte optional field, while the UDP header contains only an 8-byte protocol field. To address the issue of misaligned features arising from different packet and protocol formats, we apply a padding mechanism, where non-existent fields are filled with 0xff. For instance, fields required for TCP are padded in UDP packets, and vice versa. This alignment ensures that each byte location corresponds to the same protocol field across all packets, allowing consistent feature extraction regardless of protocol type or format. The reconstructed network and transport layer features are concatenated to form a fixed-length 128-byte header representation, denoted as $h_i$. Furthermore, this method can be extended to accommodate proprietary or other network protocols commonly found in IoT environments.

The packet payload provides complementary semantic information that is essential for understanding device interaction patterns and detecting security threats \cite{payload}. Payload data often contains textual or symbolic content, such as authentication details and device-specific information, which can offer critical insights for discriminating between specific types of malicious traffic. Given the model's requirement for consistent input format and privacy protection considerations, we apply specific processing to the hexadecimal payload of packets. Specifically, we first determine a fixed payload byte length $P$, which serves as the standard length for all packet payloads. During processing, packet payloads exceeding $P$ bytes are truncated, while those shorter than $P$ bytes undergo zero-padding to attain the fixed length of $P$ bytes. Consequently, the payload representation of each IoT traffic packet is denoted as a vector $d_i$ of length $P$ bytes.

Finally, the packet header representation $h_i$ and payload representation $d_i$ , derived from parsing each packet bytes $pkt_i$, are converted from hexadecimal to decimal, producing an aligned multimodal packet-wise representation that serves as the input for the subsequent resource-efficient neural network, as defined in Eq.(\ref{eq:1}).
\begin{equation}
\label{eq:1}
\begin{aligned}
    h_i &= [b^h_1, b^h_2, \dots, b^h_{128}], b^h_j \in [0, 255] \\
   d_i &= [b^d_1, b^d_2, \dots, b^d_P],  b^d_j \in [0, 255]
\end{aligned}
\end{equation}

The obtained aligned packet-wise representation not only enables subsequent models to effectively capture the protocol structures and payload semantic features of different protocols, but also significantly reduces the input feature dimensionality compared to other packet-level methods, which use up to 784 or 1500 bytes of features, thereby improving both performance and resource efficiency.

\begin{figure}[tbp]
    \centering
    \includegraphics[width=0.75\linewidth]{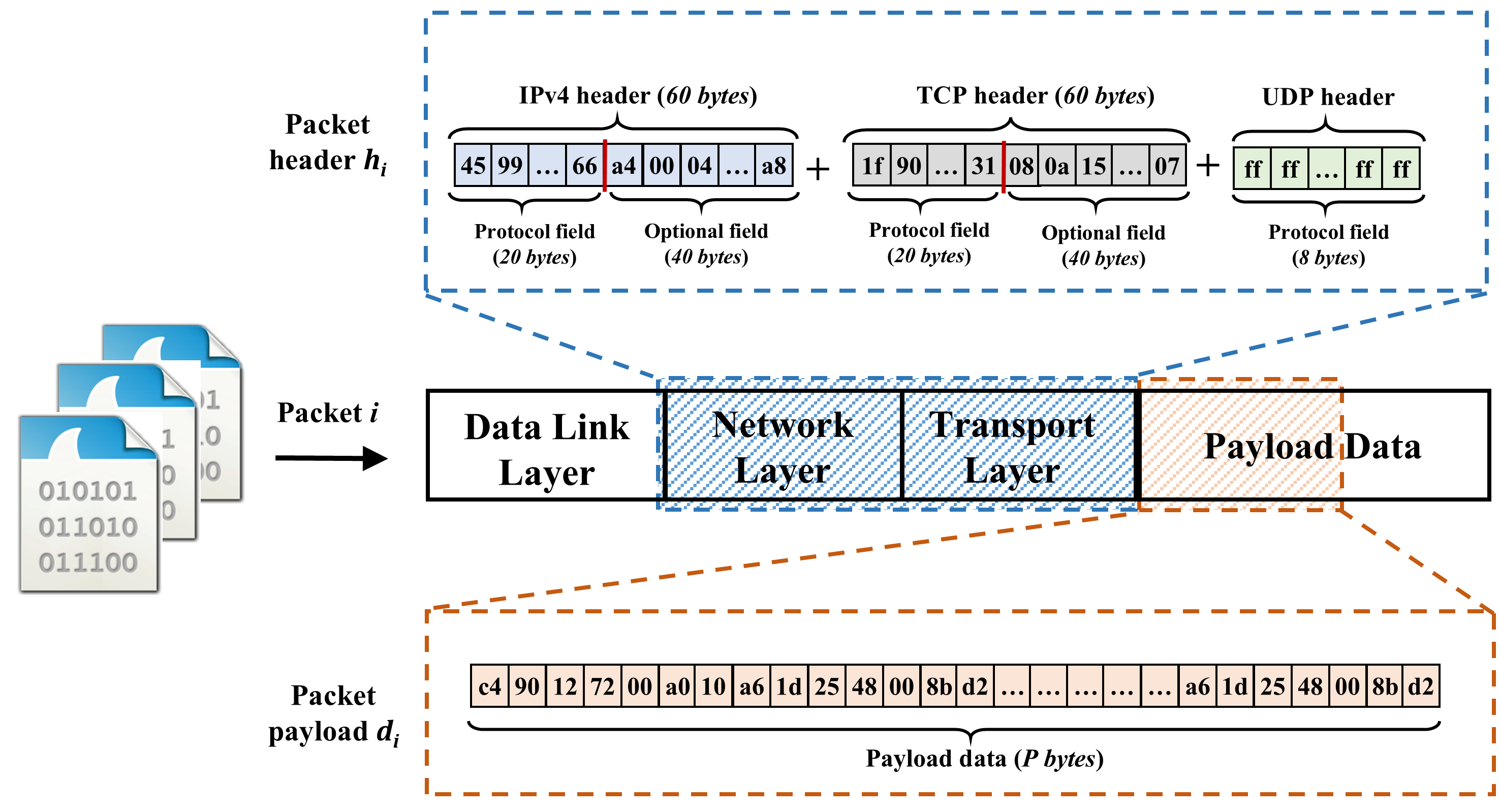}
    \caption{Illustration of the packet-wise header-payload representation.}
    \label{fig::Multimodel feature extraction}
\end{figure}

\subsection{Resource-efficient neural network}
The resource-efficient neural network aims to effectively extract and fuse multimodal traffic features from packet-wise header-payload representations using a lightweight architecture, thereby enhancing the ability to identify malicious traffic. It consists of a multimodal feature extraction module and a multimodal fusion module.

\subsubsection{Multimodal feature extraction module}

The multimodal feature extraction module is designed to further extract the most discriminative multimodal features from the constructed packet-wise header-payload representation. The module comprises two dedicated feature extractors: one for the packet header modality and another for the payload modality. These extractors are carefully tailored for their respective modality inputs, aiming to mine richer and more discriminative semantic features that spanning multiple bytes at various scales, while maintaining a lightweight architecture suitable for resource-constrained IoT environments.

Specifically, the packet header extractor focuses on mining multi-scale features from the protocol fields, capturing intricate structural patterns and variations. The payload extractor mines the semantic features embedded within the payload data, which can reveal malicious code signatures or abnormal payloads. By dedicating separate feature extractors to different modalities, this module can effectively capture the unique characteristics of both the packet header and payload modalities. The extracted multimodal features are then fed into a subsequent multimodal fusion module for further fusion.

\begin{figure}[htbp]
    \centering
    \includegraphics[width=0.7\linewidth]{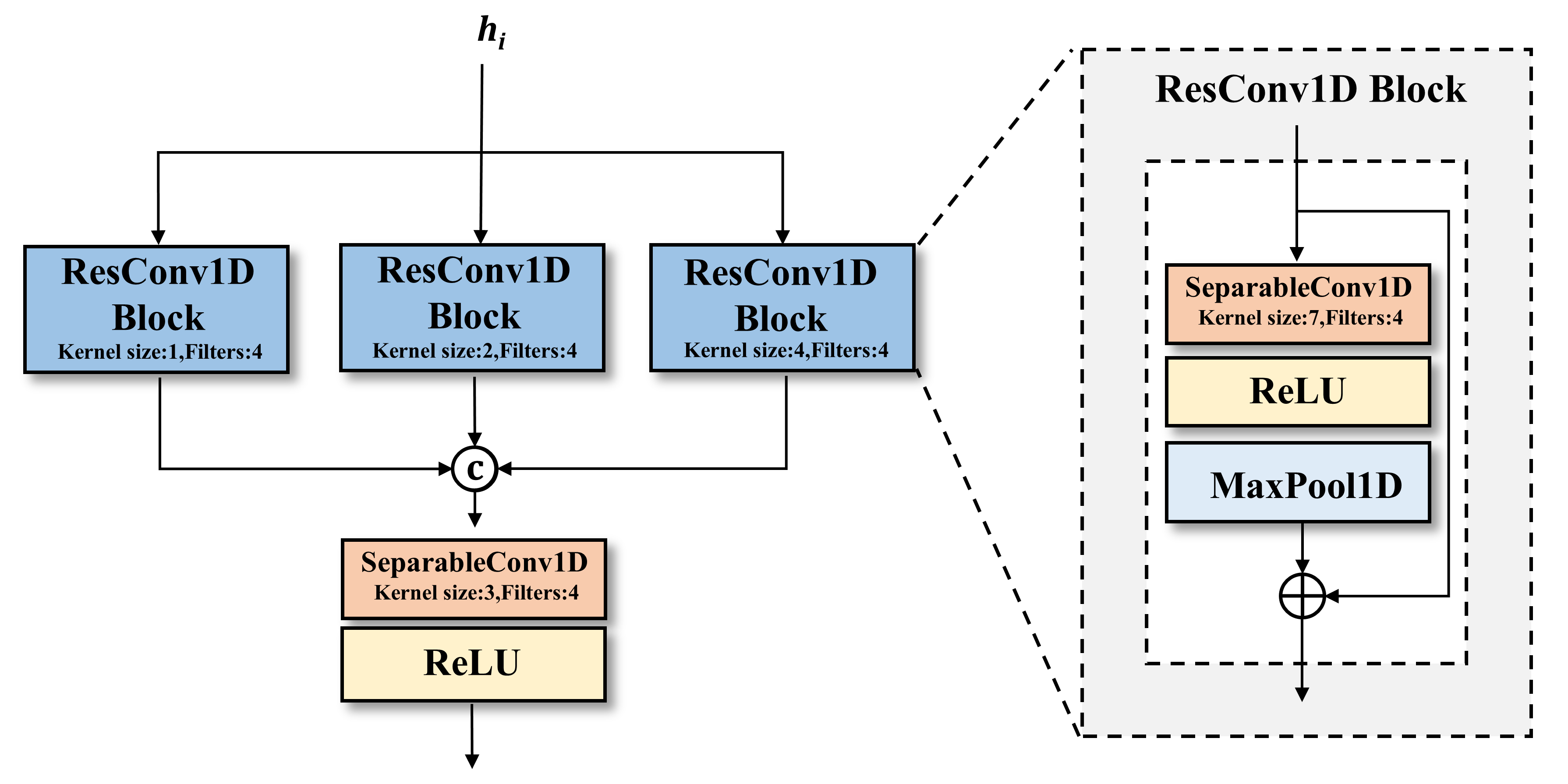}
    \caption{The structure of the header feature extractor.}
    \label{fig::header extractor}
\end{figure}

The header feature extractor employs parallel residual one-dimensional (1D) convolutional blocks with multi-scale kernel sizes to capture structural and semantic patterns within protocol fields that span multiple bytes, as shown in Fig. \ref{fig::header extractor}. Protocol fields in packet headers often exhibit diverse structural characteristics, ranging from single-byte flags (e.g., the TOS field in IPv4) to multi-byte fields (e.g., the acknowledgment number in TCP). To address this diversity, three parallel ResConv1D blocks with kernel sizes of 1, 2, and 4 are used. This multi-scale design enables the model to capture both fine-grained and coarse-grained features, which are critical for identifying subtle protocol anomalies and malicious traffic patterns. Specifically, the constructed header representation $h_i$ is simultaneously fed into three parallel one-dimensional convolutional blocks to generate three header feature maps at different scales: $m^h_1$, $m^h_2$, and $m^h_3$. Each residual 1D convolution (ResConv1D) block is composed of a 1D convolution layer, a ReLU activation function layer, a pooling layer, and a residual connection. The incorporation of the ReLU function introduces non-linearity, which significantly enhances the model's capacity for expression. The residual connection, by directly adding the input to the output of the block, forms a residual mapping. This architecture aids in more effective gradient propagation and helps to counteract the vanishing gradient issue prevalent in deep networks. The calculation process for the $k$-th ResConv1D block within the header feature extractor is as follows:
\begin{equation}
m^h_k = \text{ResConv1D} (h_i, \text{kernel size} = 2^{k-1}, \text{stride}=1)
\end{equation}

The payload feature extractor's network structure is similar to that of the header feature extractor, but the kernel sizes of the three ResConv1D blocks are set to 2, 4, and 8. This design facilitates more effective capture and processing of longer text information and its potential semantics within the payload. By inputting the constructed payload representation $d_i$, the computation process of the $k$-th ResConv1D block in the payload feature extractor is as follows:
\begin{equation}
m^p_k = \text{ResConv1D} (d_i, \text{kernel size} = 2^k, \text{stride}=1)
\end{equation}

The multi-scale features extracted from the header and payload modalities are first concatenated along the channel dimension. Subsequently, a 1D convolutional layer is applied to further refine and transform the concatenated features to generate the extracted multimodal representations $\hat{h}_i$ and $\hat{d}_i$. This transformation ensures compatibility with the input requirements of the subsequent multimodal fusion module.

Furthermore, to conserve computational and spatial resources, depthwise separable convolutions are employed instead of standard convolutions in these feature extractors. Depthwise separable convolution decomposes standard convolution into depthwise convolution and pointwise convolution, significantly reducing computational costs and model parameters. This approach enhances computational efficiency and reduces memory consumption while maintaining robust feature extraction capabilities.

\subsubsection{Multimodal fusion module}
To effectively fuse features from different modalities, the proposed multimodal fusion module employs a multi-head self-attention mechanism to dynamically focus on the most discriminative features of the extracted header and payload features to achieve the optimal fusion of complementary features from different modalities, thus further improving the classification performance of IoT-AMLHP. The module consists of $L$ identical layers, each comprising two key sub-layers: a multi-head self-attention mechanism and a feed-forward neural network, as illustrated in Fig. \ref{model}. Both sub-layers are equipped with residual connections and layer normalization (LN) to stabilize training and ensure efficient gradient propagation.

The multimodal fusion module takes as input the unified feature representation $F$ which is constructed by concatenating the extracted multimodal features $\hat{h}_i$ and $\hat{d}_i$ along the channel dimension: $F=\text{Concat}(\hat{h}_i,\hat{d}_i)=[f_1,f_2,\ldots,f_c] \in \mathbb{R}^{c \times d}$, where $c$ is the number of feature channels, and $d$ is the feature dimension. To effectively capture features from different modalities and facilitate feature fusion, $F$ is first fed into a sub-layer based on the multi-head self-attention mechanism.This sub-layer constructs queries ($Q$), keys ($K$), and values ($V$) from the input feature representation through linear projections, as defined by the following equations:  
\begin{equation}
    \begin{cases}
        Q = FW_Q\\
        K = FW_K \\
        V = FW_V
    \end{cases}
\end{equation}
where $W_Q$, $W_K$, $W_V \in \mathbb{R}^{d \times d_k}$ are learnable weight matrices, and $d_k$ is the dimension of the queries and key-value pairs for each head. Then, for each head, the attention scores are computed using scaled dot-product attention:
\begin{equation}
    \text{Attention}(Q, K, V) = \text{softmax}(\frac{QK^T}{\sqrt{d_k}})V
\end{equation}
where softmax denotes the softmax activation function, and $\frac{1}{\sqrt{d_k}}$ is a scaling factor for numerical stability. In practice, the multi-head self-attention mechanism involves multiple heads to compute the scaled dot-product attention:
\begin{equation}
    \text{MHSA}(Q, K, V) = \text{Concat}(head_1, …, head_h)W_O
\end{equation}
where $head_i =\text{Attention}(Q_i, K_i, V_i)$, and $Q_i$, $K_i$, $V_i$ are respective portions of $Q$, $K$, and $V$, and $h$ is the number of attention heads. The output of the multi-head self-attention layer is then obtained through a residual connection and layer normalization:
\begin{equation}
    F_{\text{output}} = \text{LN}(F + \text{MHSA}(Q, K, V))
\end{equation}

The second sub-layer, a feed-forward neural network (FFN), consists of two linear layers with a ReLU activation function, enabling the module to learn complex, non-linear transformations of the fused features. The output of the FFN is also combined with the input via a residual connection and normalized:
\begin{align}
    &F_{\text{final}} = \text{LN}(F_{\text{output}} + \text{FFN}(F_{\text{output}}))\\
    &\text{FFN}(F_{\text{output}}) = \text{max}(0, F_{\text{output}}W_1+b_1)W_2+b_2
\end{align}
where $W_1$, $W_2$, $b_1$, and $b_2$ are learnable parameters of the linear layers, and the ReLU activation introduces non-linearity to enhance the expressiveness of the model.

The multimodal fusion process described above is repeated for $L$ layers, allowing the model to capture deeper and more complex interactions between the header and payload features, thereby achieving optimal feature fusion. The resulting fused features are then fed into the classifier for traffic classification.
·

\subsection{Classification output}
The final component of IoT-AMLHP is a classifier, which takes the fused multimodal features as input and performs the ultimate packet classification. This classifier comprises two fully connected linear layers. The first linear layer, containing 32 neurons, performs an initial mapping and transformation of the fused features. The number of neurons in the second linear layer is set according to the specific classification task to accommodate diverse traffic classification requirements. Subsequently, the softmax function processes the classifier's output, yielding a predictive probability distribution across classes. The class with the highest probability is assigned as the predicted label for the packet. During model training, the loss function $\mathcal{L}_{\text{CE}}$ employed is the cross-entropy loss function, defined as follows:

\begin{equation}
\mathcal{L}_{\text{CE}} = -\sum_{i=1}^{N} \sum_{k=1}^{K} y_{i,k} \log (P(y=k|x_i))
\end{equation}
where $N$ represents the total number of packets, $K$ denotes the number of classes, $y_{i,k}$ is the ground truth label for packet $x_i$, and $P(y=k|x_i)$ is the predicted probability that packet $x_i$ belongs to class $k$.

\section{Performance evaluation} \label{sec::experiments}

This section presents a series of experiments conducted to validate the proposed method's effectiveness and deployability, discussed in detail in the following subsections.

\subsection{Datasets}

To evaluate our method, we conducted experiments using three public IoT traffic datasets. These datasets are commonly utilized in IoT traffic analysis due to their preservation of raw traffic files and their collection across diverse network environments, making them suitable for evaluation.

\begin{itemize}
\item{\textbf{IoT-Network-Intrusion (IoT-NI) dataset} \cite{IoT_network_intrusion_dataset} simulates network attacks in an IoT environment, employing Nmap as the primary tool. It includes a home camera and a smart speaker, which are two popular smart home appliances. This dataset meticulously captures both benign and malicious traffic. Detailed information about various benign and malicious traffic types is provided.}

\item{\textbf{ToN-IoT dataset} \cite{ToN_IoT} is created by the Canberra Network-Wide Laboratory at the University of New South Wales by simulating real-world IoT environments. It includes both benign and malicious traffic generated by devices such as smart garage doors, smart fridges, and motion-enabled lights, simulating operations in a physical IoT network. The dataset provides a labeled set of diverse abnormal IoT traffic types. For this study, we extract eight types of malicious traffic, including backdoor, DDoS, and injection attacks, in addition to benign traffic.}

\item{\textbf{MQTT-IoT-IDS2020 (MQTT-IDS) dataset} \cite{MQTT_IDS} is an intrusion detection dataset that encompasses message queuing telemetry transport (MQTT) traffic. It was generated through a simulated MQTT network environment consisting of twelve sensors, a broker, a simulated camera, and an attacker node. In this study, traffic samples containing five distinct scenarios are randomly selected to evaluate the performance of the model.
}

\end{itemize}

Furthermore, the IoT traffic samples are relabeled based on device IP, MAC address, and attacker IP. To avoid introducing biased information into the method during training, we anonymize the IP address of each packet (by replacing them with zeros). The datasets are randomly split into training, validation, and test sets following an 8:1:1 ratio. Table \ref{table::datasets1} presents the distribution of traffic types across the three datasets.

\begin{table}[htbp]
\renewcommand{\arraystretch}{1.1}
% \fontsize{10pt}{\baselineskip}
\centering
\caption{Description of the IoT-NI dataset, ToN-IoT dataset, and MQTT-IDS dataset.}
\label{table::datasets1}
\resizebox{\linewidth}{!}{
\begin{tabular}{lccc|lccc|lccc}
\toprule[1pt]
\multicolumn{4}{c|}{\textbf{IoT-NI}}                                            & \multicolumn{4}{c|}{\textbf{ToN-IoT}}                                      & \multicolumn{4}{c}{\textbf{MQTT-IDS}}                                           \\ \midrule[1pt]
\textbf{Category}      & \textbf{Label} & \textbf{\#Flows} & \textbf{\#Packets} & \textbf{Category} & \textbf{Label} & \textbf{\#Flows} & \textbf{\#Packets} & \textbf{Category}      & \textbf{Label} & \textbf{\#Flows} & \textbf{\#Packets} \\ \midrule[1pt]
Benign                 & 0              & 126              & 21523              & Benign            & 0              & 5552             & 26137              & Benign                 & 0              & 37684            & 213284             \\
Dos SYN flooding       & 1              & 59428            & 64642              & Backdoor          & 1              & 2968             & 35242              & MQTT brute-force       & 1              & 2913             & 43145              \\
Mirai ACK flooding     & 2              & 30892            & 75632              & DDos              & 2              & 10694            & 34429              & Aggressive scan        & 2              & 3939             & 4030               \\
Mirai HTTP flooding    & 3              & 163              & 1924               & Dos               & 3              & 18787            & 179056             & UDP scan               & 3              & 2253             & 2253               \\
Mirai host brute-force & 4              & 3620             & 10464              & Injection         & 4              & 17504            & 218238             & Sparta SSH brute-force & 4              & 2765             & 41475              \\
Mirai UDP flooding     & 5              & 500              & 48874              & Password          & 5              & 13106            & 114751             & \multicolumn{1}{c}{}   &                &                  &                    \\
MITM APR spoofing      & 6              & 335              & 12797              & Ransomware        & 6              & 2449             & 27694              & \multicolumn{1}{c}{}   &                &                  &                    \\
Port scan              & 7              & 15842            & 20937              & Scanning          & 7              & 15273            & 17614              & \multicolumn{1}{c}{}   &                &                  &                    \\
Operating system scan  & 8              & 417              & 1811                & Xss               & 8              & 9436             & 39578              & \multicolumn{1}{c}{}   &                &                  &                    \\ \bottomrule[1pt]
\end{tabular}
}
\end{table}

\subsection{Experimental settings}
\subsubsection{Baselines}
To demonstrate the effectiveness of our method, we introduce and reproduce seven typical state-of-the-art 
flow-level traffic classification methods as follows: % 我们选择了以下流级方法和包级方法
\begin{itemize}

\item{\textbf{1DCNN} \cite{1DCNN} leverages the first 784 bytes of each flow to construct grayscale images. These images are subsequently classified using a proposed 1D-CNN model.}

\item{\textbf{ATTLSTM} \cite{ATTLSTM} utilizes the first 1500 bytes from the initial 10 packets of each flow and integrates them into a neural network model based on an attention mechanism combined with LSTM for traffic classification.}

\item{\textbf{CNNLSTM} \cite{CNNLSTM} extracts the first 784 bytes from the initial three packets of each flow. These bytes are reshaped into $28\times28$ matrices. A CNN model is employed to extract features from each packet individually, which are then fed into an LSTM to classify the flow.}

\item{\textbf{IoT-ETEI} \cite{IoTETEI} utilizes the first 10 packets of each flow, with each packet restricted to 250 bytes. A classification model based on a CNN and BiLSTM is then applied to the flow data.}

\item{\textbf{TSCRNN} \cite{tscrnn} employs the first 15 packets of each flow, with a restriction of 1500 bytes per packet. A two-layer 1D-CNN, coupled with a BiLSTM model, is used for traffic classification.}

\item{\textbf{APPNET} \cite{APPNET} uses a bidirectional two-layer LSTM. It takes the packet length sequence as time series input for the traffic classification model, mapping the relationship between model representations and traffic labels.}

\item{\textbf{MATEC} \cite{matec} is a lightweight flow-level classification method that extracts the first 784 bytes from the first three packets of a flow, along with the positions and lengths of the packets. It utilizes a neural network combining multi-head attention and convolution to classify traffic flows.}

\end{itemize}

We also compare the proposed IoT-AMLHP with other packet-level methods:

\begin{itemize}
\item{\textbf{SAM} \cite{SAM} utilizes the initial 50 bytes of each packet as input and proposes a traffic classification approach combining a self-attention mechanism with a two-layer 1D-CNN.
}

\item{\textbf{LSTM} \cite{LSTM} uses the first 784 bytes of each packet. It employs a three-layer LSTM to identify the packet’s category.}

\item{\textbf{Deeppacket} \cite{Deeppacket} is a packet-level method that uses the first 1500 bytes of each packet. It classifies the packets using a neural network composed of two consecutive convolutional layers.}

\item{\textbf{MISCNN+} \cite{MISCNN+} is a lightweight method that processes the first 784 bytes of each packet, reshapes them into various formats, and feeds them into a residual convolutional network consisting of stacked residual blocks for classification.}

\end{itemize}

\subsubsection{Implementation details}

For the proposed method, the number of heads $h$ in the multimodal fusion module is set to 8. Subsequently, we perform hyperparameter tuning for the number of layers $L$ in the multimodal fusion module, exploring values within the range of $\{1,2,4,8\}$, and the payload bytes length $P$, spanning the range of $\{16,64,128\}$. Additional information about the hyperparameter tuning process will be discussed in the following subsection.

Experiments were conducted on a Windows 11 system with an NVIDIA RTX 3090 GPU and 24GB RAM. The foundation of our software platform relies on Python 3.9 and PyTorch 1.11.0. We utilize the SplitCap tool and Scapy during the preprocessing stage of the traffic data.

\subsubsection{Evaluation metrics}
To evaluate the classification performance of our proposed method, we utilize four evaluation metrics: Accuracy (ACC), Precision (PR), Recall (RC), and F1-score (F1). A higher value for these metrics indicates better performance. The calculation rules for these metrics are as follows:
\begin{align}
    \text{Accuracy (ACC)} &= \frac{\text{TP}+\text{TN}}{\text{TP}+\text{FP}+\text{TN}+\text{FN}} \\
    \text{Precision (PR)} &= \frac{\text{TP}}{\text{TP}+\text{FP}} \\
    \text{Recall (RC)} &= \frac{\text{TP}}{\text{TP}+\text{FN}}  \\
    \text{F1-score (F1)} &= 2 \cdot \frac{\text{Precision} \cdot \text{Recall}}{\text{Precision} + \text{Recall}}
\end{align}
where $\text{TP}$, $\text{TN}$, $\text{FP}$, and $\text{FN}$ denote the true positive, true negative, false positive, and false negative, respectively.

Given that malicious traffic classification typically involves multiple categories, we employ macro average method to determine the mean value of each evaluation metric across all classes.
\begin{align}
    \text{Macro-}(\text{PR}, \text{RC}, \text{F1}) &= \frac{\sum_{k \in \text{classes}} (\text{PR}, \text{RC}, \text{F1})_k}{N}
\end{align}
where $(\text{PR}, \text{RC}, \text{F1})_k$ denotes the $k$-th value classification, and $N$ represents the number of classifications. In this experiment, the accuracy for the prediction of all classification accuracy, the other evaluation metrics belong to the macro average.

Furthermore, Floating-Point Operations (FLOPs), the number of parameters (Params), and model size are used to evaluate the computational and storage overhead of the methods.

\begin{table}[htbp]
\renewcommand{\arraystretch}{1.1}
\centering
\caption{Hyperparameters tuning for the proposed IoT-AMLHP.}
\label{tab:hyperparameters}
\resizebox{0.6\linewidth}{!}{
\begin{tabular}{lll}
\toprule[1pt]
\textbf{Hyperparameters} & \textbf{Search range} & \textbf{Final selection} \\ \midrule[1pt]
Training epoch            & [5,10,20,30]         & 10                      \\ 
Training batch size      & [64,128,256,512]          & 128                      \\ 
Optimizer                 & [SGD,Adam,AdaGrad]   & Adam                     \\
Learning Rate            & [1E-4,1E-3,1E-2,1E-1] & 1.00E-02                 \\ 
Dropout Rate             & [0.1,0.2,0.3]         & 0.1                      \\ 

$L$                      & [1,2,4,8]       & 2                      \\ 
$P$                      & [16,64,128]       & 64                       \\ \bottomrule[1pt]
\end{tabular}
}
\end{table}

\subsection{Hyperparameter tuning}
In this study, to ensure optimal performance of the proposed IoT-AMLHP, we performed hyperparameter tuning on the IoT-NI dataset, striking a balance between classification performance and computational complexity. During the training phase, we observed the variations in performance metrics by adjusting the values of hyperparameters to determine their optimal selection. Table \ref{tab:hyperparameters} presents the determined hyperparameters and their respective search ranges.

\textbf{\begin{figure}[htbp]
  \centering
    \begin{minipage}[b]{\linewidth}
    \centering
      \subfigure[Metrics]{
        \includegraphics[width=0.43\linewidth]{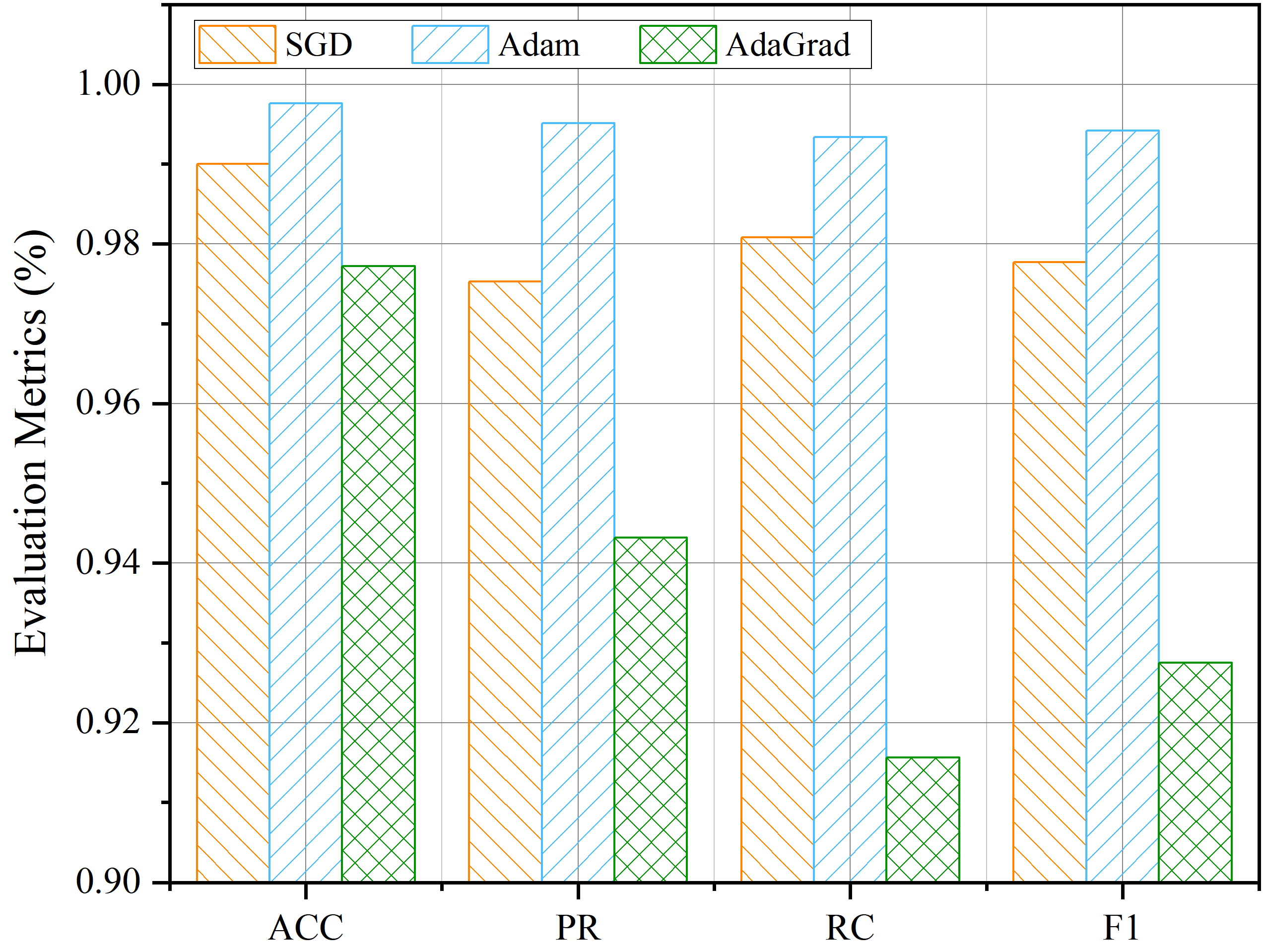}
      }
      \subfigure[Loss]{
        \includegraphics[width=0.41\linewidth]{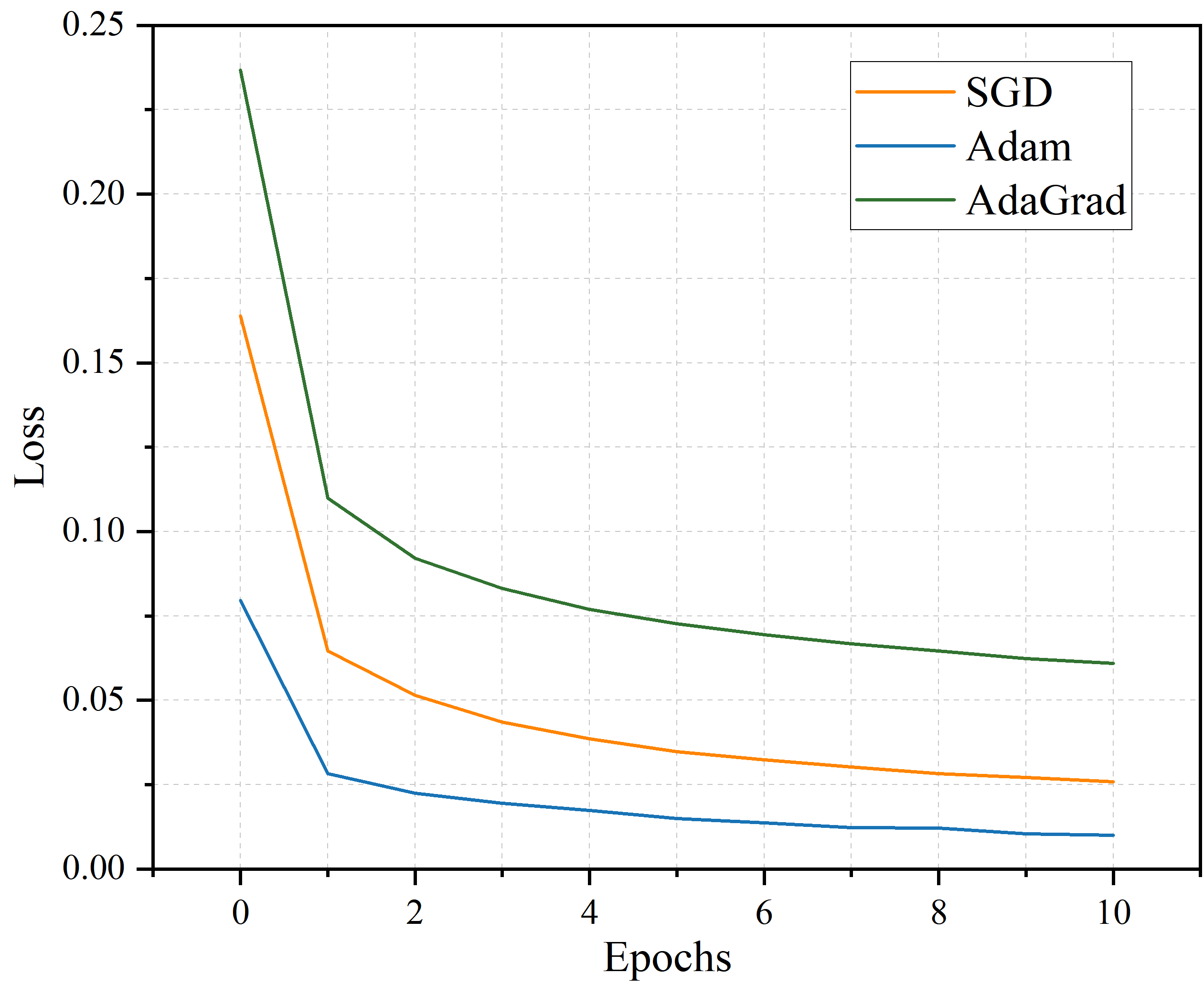}
      }
    \end{minipage}
    \caption{The hyperparameter tuning of IoT-AMLHP under different optimizers. (a) Comparison of evaluation metics. (b) Loss convergence over train epochs. }
    \label{fig::hyper_optimizer}
\end{figure}}

To evaluate the impact of different optimization algorithms, we conducted experiments using three widely used optimizers: Stochastic Gradient Descent (SGD), Adam, and AdaGrad. As shown in Fig. \ref{fig::hyper_optimizer}, the Adam optimizer outperformed the other two optimizers across all evaluation metrics. Specifically, Adam achieved the highest performance metrics, owing to its adaptive learning rate mechanism, which stabilizes training and enables faster convergence. In comparison, SGD demonstrated slower convergence and required more epochs to achieve acceptable results, while AdaGrad plateaued early, limiting its performance gains. Based on these results, Adam was selected as the final optimizer for IoT-AMLHP, balancing classification accuracy and computational efficiency.

Batch size, as an important hyperparameter, also affects the classification performance of the IoT-AMLHP framework. We evaluated batch sizes of 64, 128, 256, and 512 on the IoT-NI dataset to determine the optimal setting, as illustrated in Fig. \ref{fig::hyper_b_l} (a). Smaller batch sizes (e.g., 64 and 128) exhibited superior performance, while larger batch sizes (e.g., 512) led to a slight decrease in accuracy and F1-score. The reduction in accuracy and F1-score for larger batch sizes may be attributed to fewer updates per epoch, limiting the model's ability to effectively generalize. A batch size of 128 achieved the optimal performance metrics while maintaining an effective balance between memory usage and training speed. Consequently, all subsequent experiments were conducted with a batch size of 128.

We further evaluated the impact of different learning rates on the IoT-AMLHP framework, as shown in Fig. \ref{fig::hyper_b_l} (b). The experimental results indicate that a learning rate of 1E-1 caused instability during training, preventing the model from converging and resulting in poor performance. On the other hand, smaller learning rates, such as 1E-3 and 1E-4, demonstrated stable convergence but suffered from slower convergence speed and reduced training efficiency. Considering both training stability and convergence speed, the learning rate of 1E-2 achieved the best performance. Therefore, 1E-2 was selected as the final learning rate to balance performance and efficiency.

\textbf{\begin{figure}[htbp]
  \centering
    \begin{minipage}[b]{\linewidth}
    \centering
      \subfigure[Batch size]{
        \includegraphics[width=0.43\linewidth]{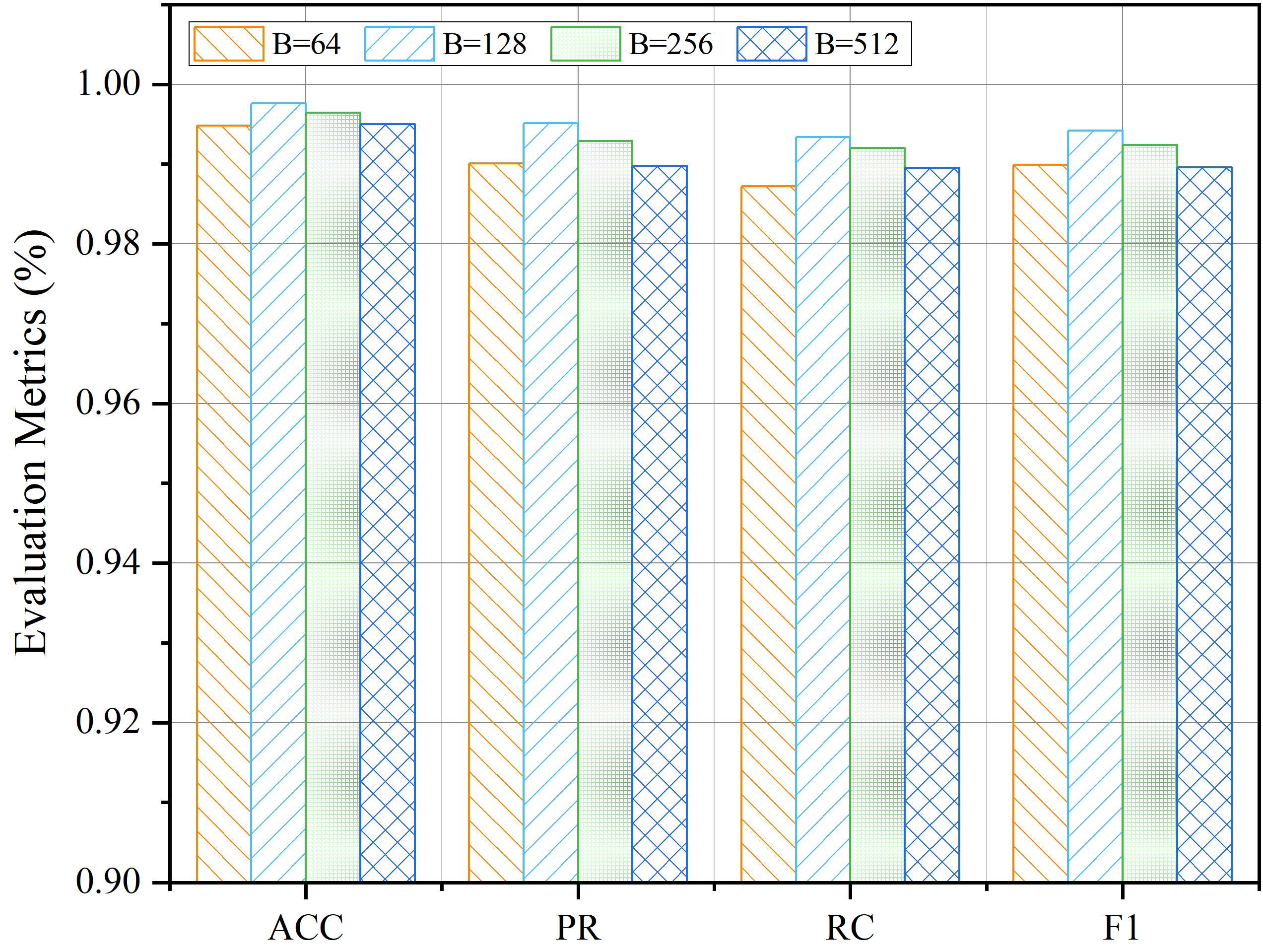}
      }
      \subfigure[Learning rate]{
        \includegraphics[width=0.43\linewidth]{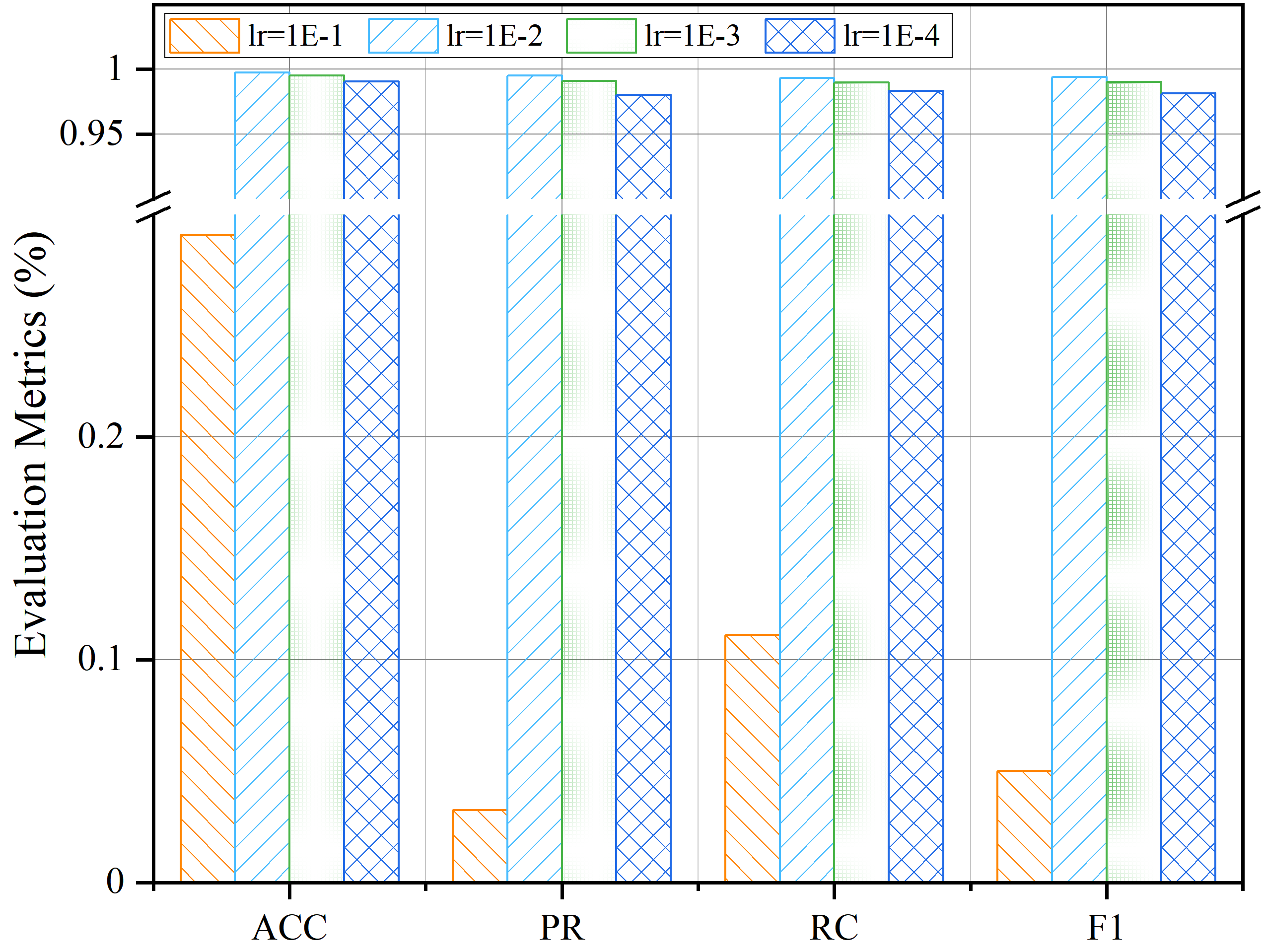}
      }
    \end{minipage}
    \caption{The hyperparameter tuning of IoT-AMLHP under different Batch sizes $B$ and learning rates $lr$. (a) Impact of different batch sizes on classification performance. (b) Impact of different learning rate on classification performance. }
    \label{fig::hyper_b_l}
\end{figure}}

\begin{figure}[htbp]
  \centering
    \begin{minipage}[b]{\linewidth}
    \centering
      \subfigure[$P$]{
        \includegraphics[width=0.43\linewidth]{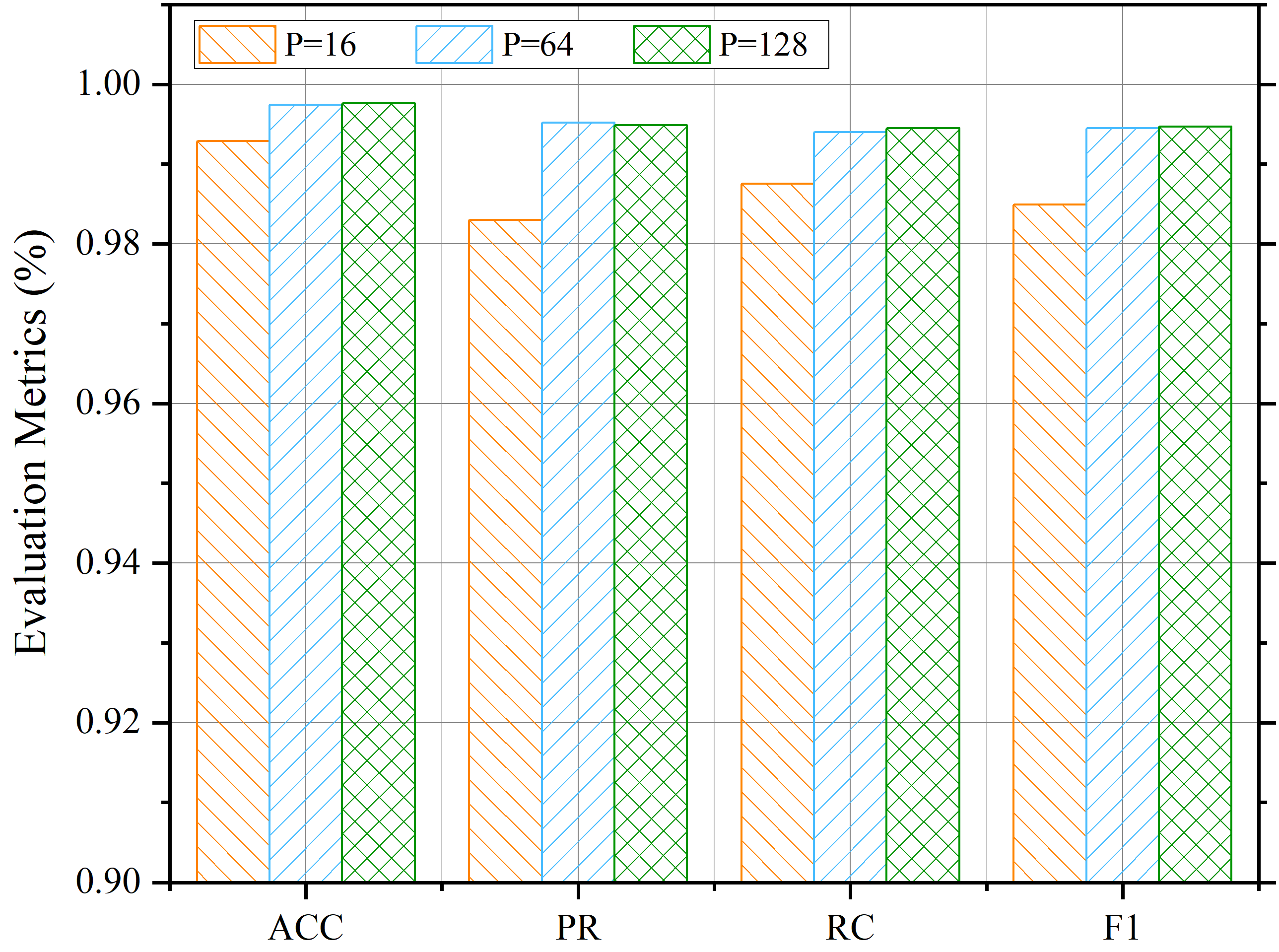}
      }
      \subfigure[$L$]{
        \includegraphics[width=0.47\linewidth]{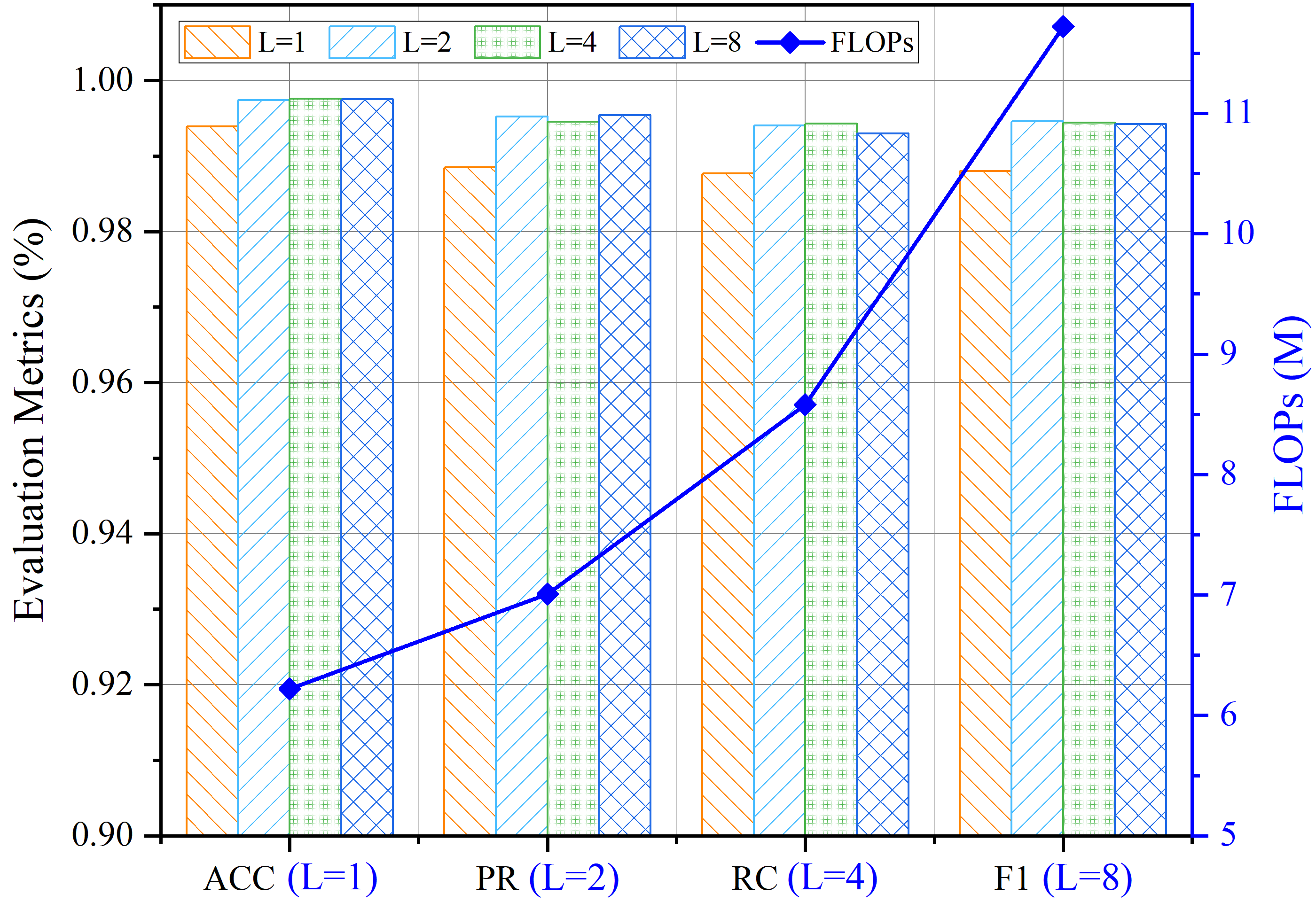}
      }
    \end{minipage}
    \caption{The hyperparameter tuning of IoT-AMLHP under different payload lengths $P$ and the number of layers $L$. (a) Impact of different $P$ values on classification performance. (b) Impact of varying $L$ on classification performance and computational complexity. }
    \label{fig::hyper}
\end{figure}

Since one of the modalities in IoT-AMLHP utilizes payload bytes for classification, this may raise privacy concerns, as payloads could potentially contain sensitive information about users and devices. Therefore, it is crucial to select an appropriate payload byte length $P$, to balance the risk of information leakage and classification performance. Specifically, a larger $P$ may lead to the exposure of sensitive data, while a smaller $P$ could compromise classification accuracy due to insufficient information. We evaluated the impact of setting $P$ to 16, 64, and 128 on the performance of the IoT-AMLHP, as depicted in Fig. \ref{fig::hyper} (a). When $P$ is larger, better classification performance can be achieved, particularly when increasing $P$ from 16 to 64. However, further increasing $P$ does not significantly improve the classification performance. Therefore, we set $P=64$, which not only ensures robust classification performance but also adequately controls privacy risks.

Additionally, the number of layers $L$ in the multimodal fusion module affects both the classification performance and the computational complexity. We investigated the impact of varying $L$ on both classification performance and model complexity. Experimental results demonstrate that increasing $L$ from 1 to 2 significantly improves performance. However, further increments in $L$ have minimal gains in classification and a noticeable increase in FLOPs, as depicted in Fig. \ref{fig::hyper} (b). Hence, considering the trade-off between classification performance and computational overhead, the number of layers $L$ is set to 2 for subsequent experiments.

\subsection{Ablation study}
This subsection conducts an ablation study to evaluate the impacts of different modalities and the effectiveness of the multimodal fusion module in the proposed IoT-AMLHP. We design the following three variant models:
\begin{itemize}
\item{\textbf{Variant 1:} This variant removes the packet header modality. It masks the packet header representation (by replacing them with zeros) before inputting them into the model for classification.}
\item{\textbf{Variant 2:} This variant removes the packet payload modality. It masks the packet payload representation before inputting them into the model for classification.}
\item{\textbf{Variant 3:} This variant removes the multimodal fusion module and replaces it with a fully connected layer for traffic classification.}
\end{itemize}

\begin{figure}[tbp]
  \centering
    \begin{minipage}[b]{\linewidth}
    \centering
      \subfigure[IoT-NI]{
        \includegraphics[width=0.32\linewidth]{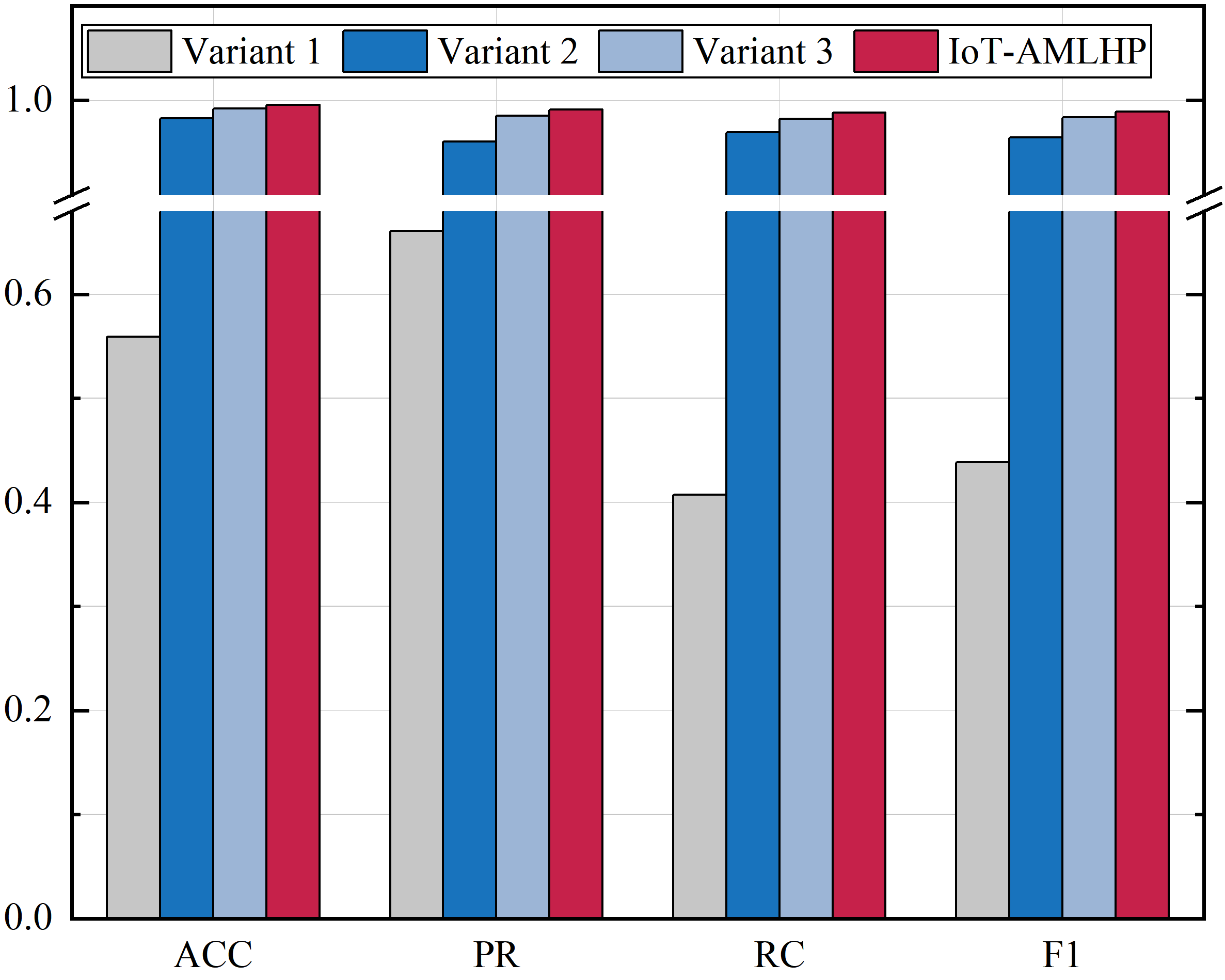}
      }
        \subfigure[ToN-IoT]{
        \includegraphics[width=0.31\linewidth]{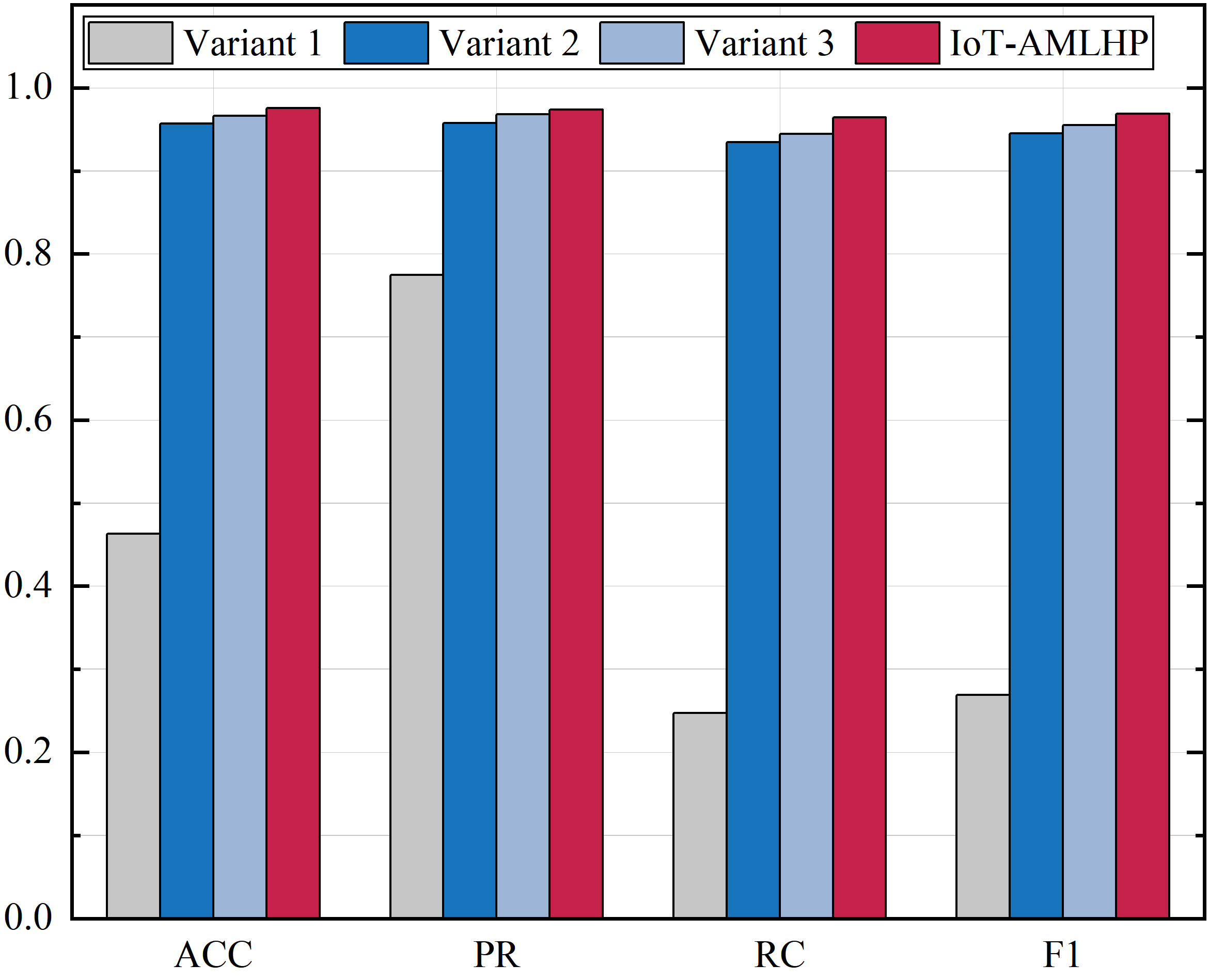}
      }
      \subfigure[MQTT-IDS]{
        \includegraphics[width=0.32\linewidth]{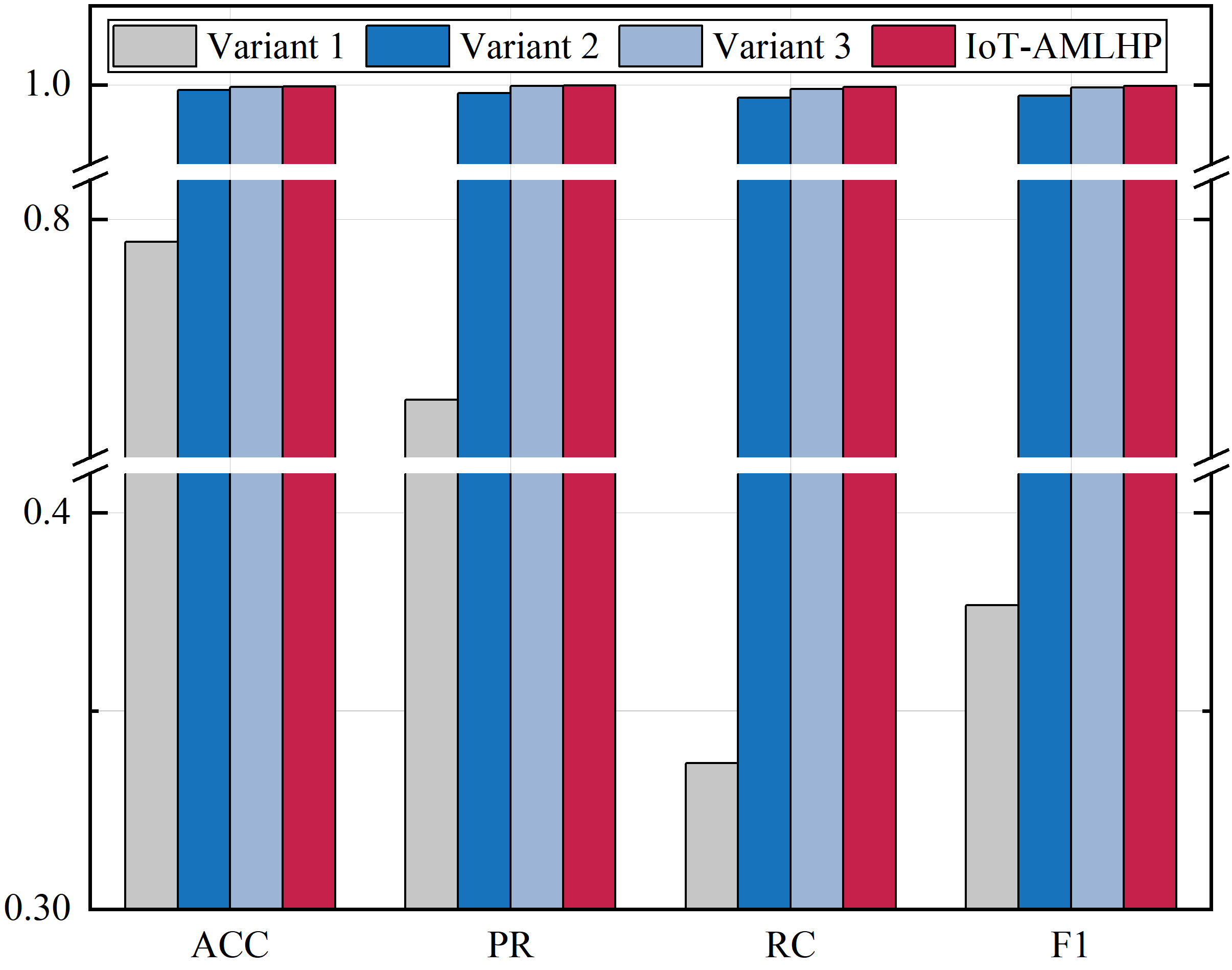}
      }
    \end{minipage}
    \caption{Ablation study of IoT-AMLHP on IoT-NI, ToN-IoT and MQTT-IDS datasets.}
    \label{fig::variant}
\end{figure}

The ablation study results, presented in Table \ref{Table::ablation_1} and Fig. \ref{fig::variant}, provide a comprehensive evaluation of each modality's contribution and the effectiveness of the multimodal fusion module in IoT-AMLHP.

The packet header contains protocol-level features that are critical for identifying structural anomalies and traffic patterns. Removing the header modality (Variant 1) led to a noticeable performance drop, with accuracy decreasing from 0.9976 to 0.5592 on IoT-NI and from 0.9996 to 0.7944 on MQTT-IDS. These results underscore the packet header's role in capturing protocol-specific irregularities, such as unexpected flag combinations or malformed packets, which are often indicative of malicious behavior.

The packet payload provides rich semantic information, crucial for identifying the content and intent of the communication. The results indicate a  performance degradation when the payload modality is removed (Variant 2), with F1-score dropping from 0.9942 to 0.9804 on IoT-NI and from 0.9691 to 0.9452 on ToN-IoT. Although the model's performance does not decline substantially when remove payload features, this does not diminish the importance of the payload modality. On the contrary, the payload provides critical insights for detecting specific types of malicious traffic.

Furthermore, the ablation experiment results of Variant 3 demonstrate the crucial role of the multimodal fusion module in identifying malicious IoT traffic. When the multimodal fusion module is replaced by a fully connected layer, the F1-score declines across all three datasets, with a decrease from 0.9691 to 0.9555 on the ToN-IoT dataset. This performance degradation stems from our fusion module's adoption of a multi-head self-attention mechanism, which effectively fuses features from different modalities, thereby enhancing the model's ability to accurately and robustly identify malicious traffic.

The ablation study validates the rationale and effectiveness of the various design components employed in the proposed method. Each part of the model contributes to the overall performance, underscoring the significance of leveraging multimodal information and effective feature fusion for accurate IoT malicious traffic identification.

\begin{table*}[tbp]
\renewcommand{\arraystretch}{1.1}
\centering
\caption{Ablation study of IoT-AMLHP on IoT-NI, ToN-IoT and MQTT-IDS datasets. Bold numbers is the best.}
\label{Table::ablation_1}
\resizebox{0.99\linewidth}{!}{
\begin{tabular}{ll|cccc|cccc|cccc}
\toprule[1pt]
\multicolumn{2}{c|}{Dataset} & \multicolumn{4}{c|}{IoT-NI} & \multicolumn{4}{c|}{ToN-IoT} & \multicolumn{4}{c}{MQTT-IDS} \\ \midrule[1pt]
\multicolumn{2}{c|}{Methods} & ACC & PR & RC & F1 & ACC & PR & RC & F1 & ACC & PR & RC & F1 \\ \midrule[1pt]
                            & Variant 1 & 0.5592 & 0.6611 & 0.4071 & 0.4384 & 0.4692 & 0.7750 & 0.2471 & 0.2688 & 0.7944 & 0.7546 & 0.3369 & 0.3767 \\
                            & Variant 2 & 0.9906 & 0.9782 & 0.9831 & 0.9804 & 0.9570 & 0.9576 & 0.9344 & 0.9452 & 0.9987 & 0.9979 & 0.9967 & 0.9973 \\
                            & Variant 3 & 0.9958 & 0.9919 & 0.9902 & 0.9910 & 0.9664 & 0.9681 & 0.9449 & 0.9555 & 0.9995 & 0.9997 & 0.9989 & 0.9993 \\ \cmidrule{2-14}
                            & IoT-AMLHP & \textbf{0.9976} & \textbf{0.9951} & \textbf{0.9934} & \textbf{0.9942} & \textbf{0.9758} & \textbf{0.9739} & \textbf{0.9645} & \textbf{0.9691} & \textbf{0.9996} & \textbf{0.9999} & \textbf{0.9995} & \textbf{0.9997} \\
\bottomrule[1pt]
\end{tabular}
}
\end{table*}

\subsection{Performance analysis}
\subsubsection{Classification performance analysis}

This section comprehensively assesses the effectiveness of the IoT-AMLHP for the IoT malicious traffic classification task. We compare its performance across two datasets against several baseline methods. For a fair comparison with other flow-level methods, we only utilize the first packet of each flow as input for IoT-AMLHP. Table \ref{Table:Performance} presents a detailed comparison of the classification performance of the proposed IoT-AMLHP against other methods across different datasets.

It can be observed that IoT-AMLHP achieves nearly state-of-the-art performance compared to flow-level baseline methods. Specifically, on the ToN-IoT dataset, IoT-AMLHP outperforms other baseline methods in all performance metrics. On the IoT-NI dataset, IoT-AMLHP outperforms other methods in terms of ACC, RC, and F1, while its PR is slightly lower than the TSCRNN. Similarly, on the MQTT-IDS dataset, IoT-AMLHP surpasses other flow-level baseline methods in ACC and F1, and exhibits strong competitiveness in PR and RC. The lower performance in certain metrics compared to other baselines is understandable, as flow-level methods typically extract richer features from multiple packets within a flow, whereas IoT-AMLHP extracts limited traffic features from individual packets. Nevertheless, IoT-AMLHP maintains balanced performance across metrics, as evidenced by its higher F1-score, which effectively integrates PR and RC, offering a more holistic evaluation of classification capability.

Compared to packet-level baseline methods, IoT-AMLHP consistently achieves superior performance, with an average accuracy exceeding 99\% across all datasets. it achieves notable gains over lightweight models such as MISCNN+, with an F1-score improvement of 1.98\% on IoT-NI and 2.70\% on ToN-IoT. These improvements underscore the effectiveness of the proposed multimodal representation and fusion mechanisms, which enable IoT-AMLHP to extract richer semantic features from both packet headers and payloads to achieve superior malicious traffic identification performance.

\begin{table*}[htbp]
\renewcommand{\arraystretch}{1.1}
\centering
\caption{Comparison of different methods on IoT-NI, ToN-IoT, and MQTT-IDS datasets. Bold numbers is the best.}
\label{Table:Performance}
\resizebox{0.99\linewidth}{!}{
\begin{tabular}{ll|cccc|cccc|cccc}
\toprule[1pt]
\multicolumn{2}{c|}{Dataset} & \multicolumn{4}{c|}{IoT-NI} & \multicolumn{4}{c|}{ToN-IoT} & \multicolumn{4}{c}{MQTT-IDS} \\ \midrule[1pt]
\multicolumn{2}{c|}{Methods} & ACC & PR & RC & F1 & ACC  & PR & RC & F1 & ACC & PR & RC & F1 \\ \midrule[1pt]
\multirow{8}{*}{Flow-level} & 1DCNN & 0.9991 & 0.9697 & 0.9525 & 0.9596 & 0.9905 & 0.9891 & 0.9886 & 0.9887 & 0.9986 & 0.9967 & \textbf{0.9974} & 0.9971 \\
                            & ATTLSTM & 0.9979 & 0.8444 & 0.8748 & 0.8585 & 0.9920 & 0.9850 & 0.9849 & 0.9849 & 0.9978 & 0.9976 & 0.9932 & 0.9954 \\
                            & CNNLSTM & 0.9992 & 0.9785 & 0.9615 & 0.9685 & 0.9943 & 0.9931 & 0.9923 & 0.9927 & 0.9980 & 0.9958 & 0.9951 & 0.9954 \\
                            % & MISCNN & 0.9973 & 0.9716 & 0.9155 & 0.9258 & {\color{blue}0.9908} & {\color{blue}0.9862} & {\color{blue}0.9844} & {\color{blue}0.9851} & 0.9984 & 0.9986 & 0.9947 & 0.9966 \\
                            & IoT-ETEI & 0.9991 & 0.9718 & 0.9480 & 0.9564 & 0.9936 & 0.9866 & 0.9875 & 0.9868 & 0.9980 & 0.9990 & 0.9924 & 0.9957 \\
                            & TSCRNN & 0.9993 & \textbf{0.9885} & 0.9642 & 0.9748 & 0.9955 & 0.9925 & 0.9928 & 0.9926 & 0.9960 & \textbf{0.9990} & 0.9877 & 0.9931 \\
                            & APPNET & 0.8193 & 0.8812 & 0.8108 & 0.8368 & 0.9711 & 0.9056 & 0.9003 & 0.8991 & 0.9966 & 0.9984 & 0.9824 & 0.9900 \\
                            & MATEC & 0.9954 & 0.7680 & 0.7613 & 0.7584 & 0.9898 & 0.9837 & 0.9850 & 0.9844 & 0.9939 & 0.9840 & 0.9882 & 0.9860 \\ \cmidrule{2-14}
                            & IoT-AMLHP & \textbf{0.9996} & 0.9788 & \textbf{0.9738} & \textbf{0.9762} & \textbf{0.9962} & \textbf{0.9958} & \textbf{0.9950} & \textbf{0.9954} & \textbf{0.9986} & 0.9978 & 0.9973 & \textbf{0.9975} \\ \midrule[0.5pt]
\multirow{5}{*}{Packet-level} & SAM & 0.9955 & 0.9933 & 0.9902 & 0.9915 & 0.9736 & 0.9705 & 0.9586 & 0.9638 & 0.9994 & 0.9977 & 0.9963 & 0.9969 \\
                              & LSTM & 0.9798 & 0.9591 & 0.9595 & 0.9590 & 0.9384 & 0.9321 & 0.9013 & 0.9154 & 0.9993 & 0.9990 & 0.9953 & 0.9971 \\
                              & DeepPacket & 0.9919 & 0.9792 & 0.9849 & 0.9214 & 0.9477 & 0.9378 & 0.9296 & 0.9325 & 0.9990 & 0.9993 & 0.9957 & 0.9975 \\
                              & MISCNN+ & 0.9896 & 0.9705 & 0.9793 & 0.9744 & 0.9528 & 0.9494 & 0.9369 & 0.9421 & 0.9977 & 0.9982 & 0.9978 & 0.9980 \\ \cmidrule{2-14}
                              & IoT-AMLHP & \textbf{0.9976} & \textbf{0.9951} & \textbf{0.9934} & \textbf{0.9942} & \textbf{0.9758} & \textbf{0.9739} & \textbf{0.9645} & \textbf{0.9691} & \textbf{0.9996} & \textbf{0.9999} & \textbf{0.9995} & \textbf{0.9997} \\
\bottomrule[1pt]
\end{tabular}
}
\end{table*}

\begin{figure}[htbp]
  \centering
    \begin{minipage}[b]{\linewidth}
    \centering
      \subfigure[IoT-NI]{
        \includegraphics[width=0.32\linewidth]{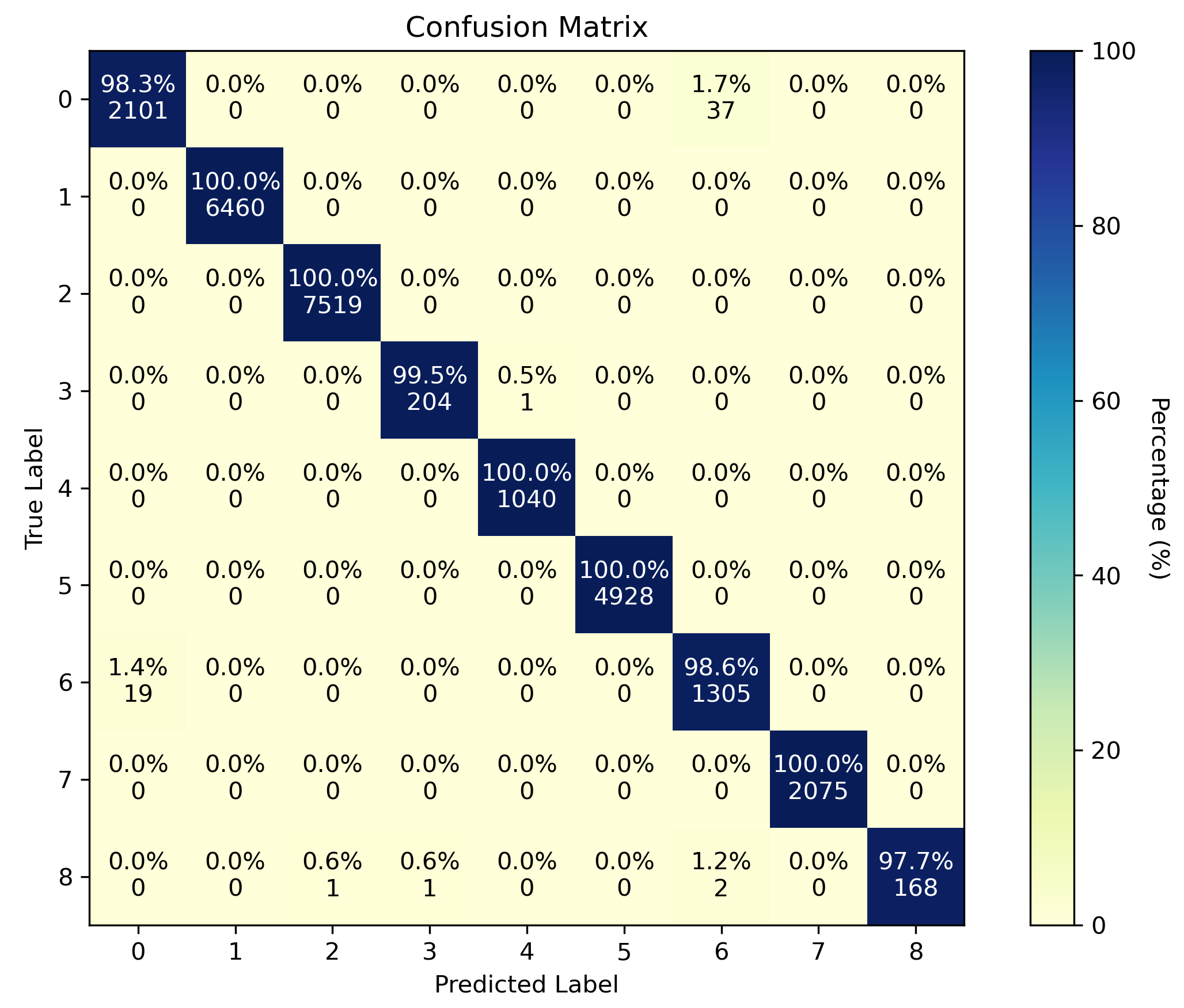}
      }
        \subfigure[ToN-IoT]{
        \includegraphics[width=0.32\linewidth]{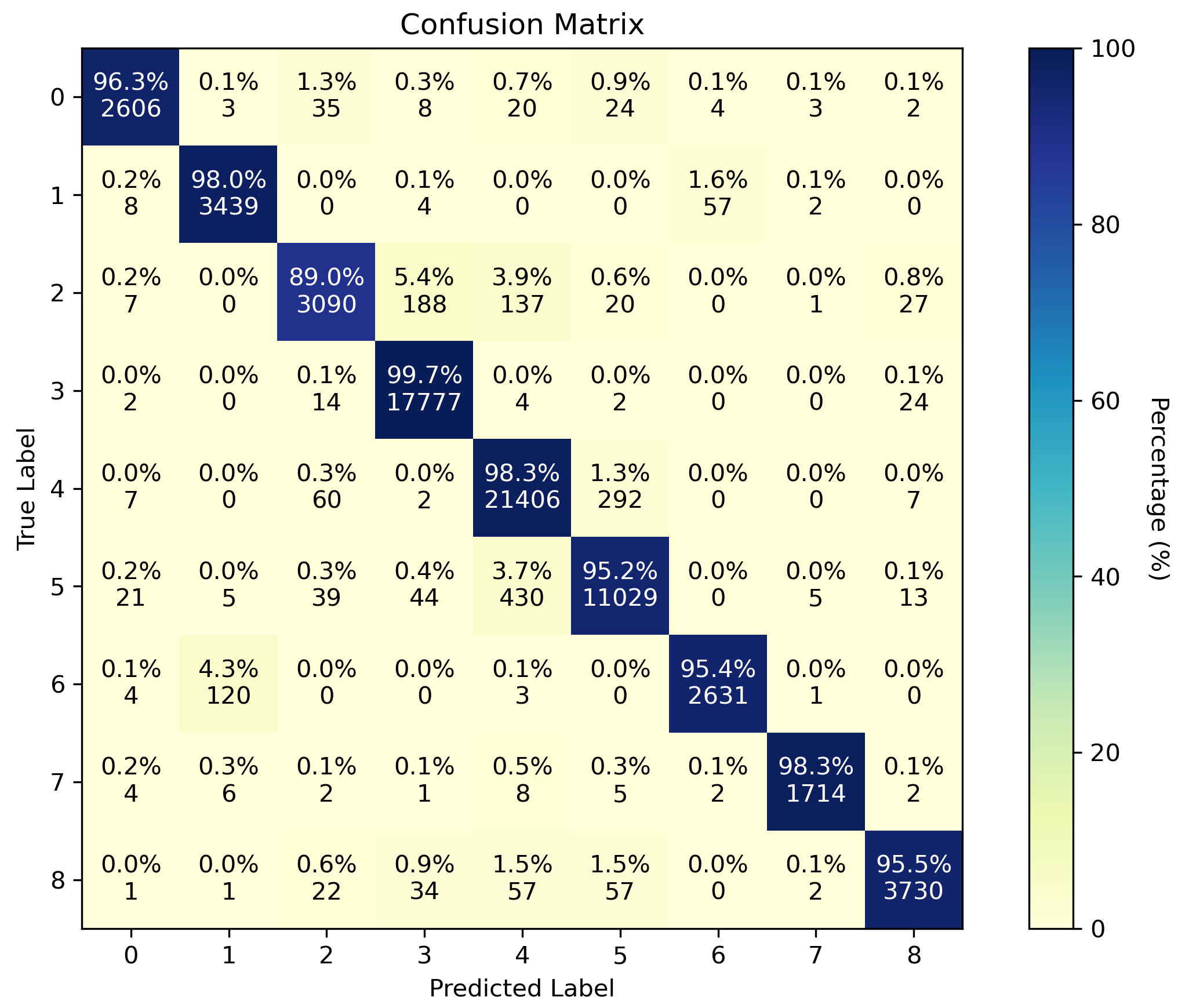}
      }
      \subfigure[MQTT-IDS]{
        \includegraphics[width=0.32\linewidth]{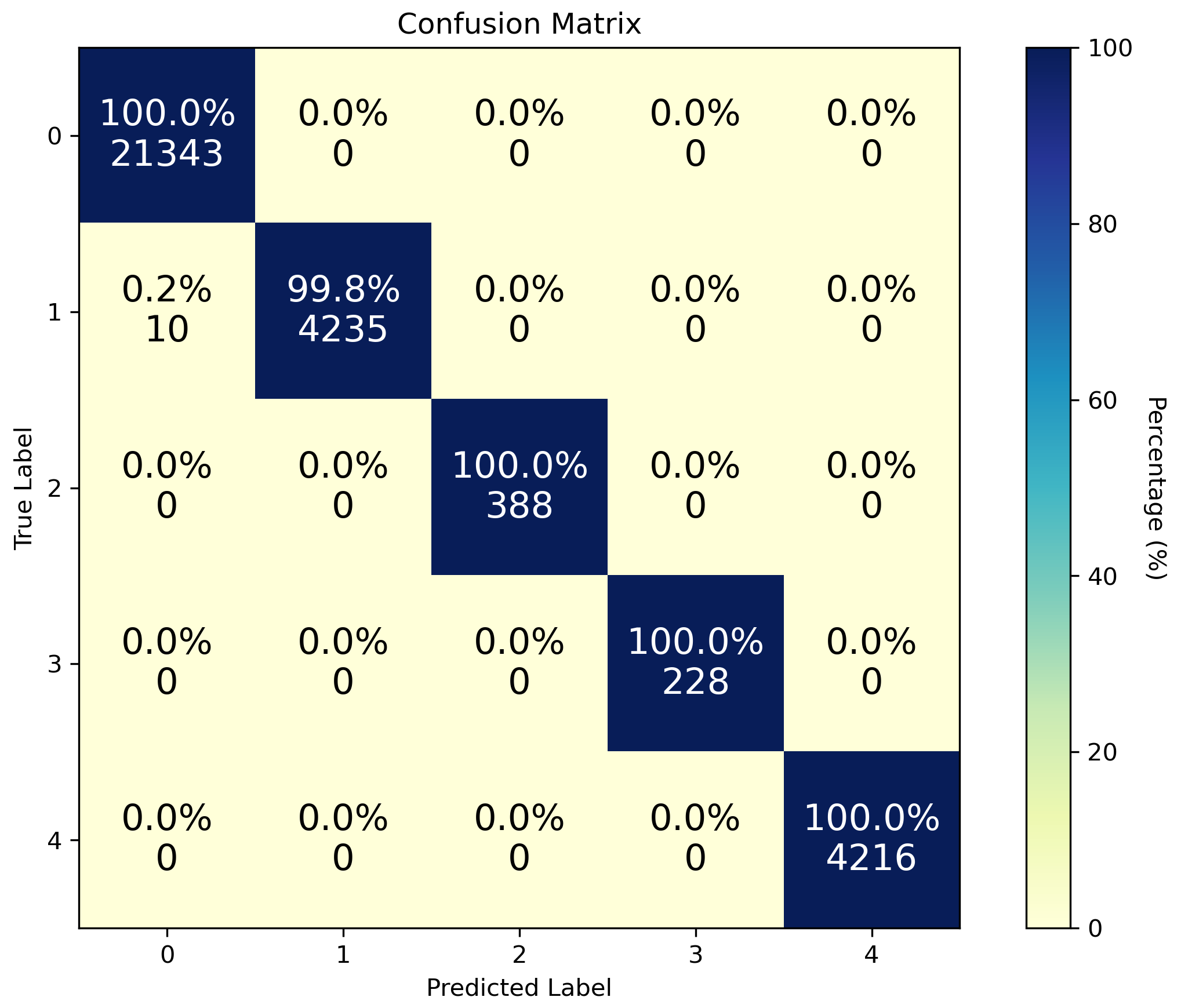}
      }
    \end{minipage}
    \caption{Confusion matrices of the proposed IoT-AMLHP. (a) Confusion matrix on the IoT-NI dataset. (b) Confusion matrix on the ToN-IoT dataset. (c)  Confusion matrix on the MQTT-IDS dataset.}
    \label{fig::confusionmatrix}
\end{figure}

To present a clear and detailed representation of the classification performance of the proposed IoT-AMLHP, Fig. \ref{fig::confusionmatrix} presents heatmaps of the confusion matrices for the three datasets. It can be clearly observed that a significant portion of the confusion matrix elements are aligned along the diagonal, represented by varying shades of blue. This feature indicates that the model achieves significant classification accuracy on all datasets, correctly classifying the vast majority of samples. Therefore, based on the quantitative evaluation metrics presented in Table \ref{Table:Performance} and the visual representation of the confusion matrices in Fig. \ref{fig::confusionmatrix}, it can be concluded that IoT-AMLHP demonstrates high efficiency and superior performance in the IoT malicious traffic classification.

\subsubsection{Analysis of computational and storage overhead}

The proposed IoT-AMLHP model is designed to achieve superior traffic classification performance while maintaining low computational and storage overhead. Metrics such as FLOPs, the number of parameters, and model size are crucial in evaluating the resource efficiency of machine learning models in IoT applications. FLOPs represent the computational workload of the model; the lower the FLOPs, the lower the computational demand, which helps save computational resources \cite{Flops}. This is a critical factor for low-power IoT devices. Furthermore, the number of parameters and model size determine the storage requirements, directly impacting the model's deployability on hardware with limited storage resources \cite{Size_1,Size_2}.

To evaluate the computational and storage overhead of IoT-AMLHP, we selected several deep learning models with comparable classification performance on the IoT-NI dataset from the baselines. As shown in Table \ref{Table::Lightweight}, IoT-AMLHP demonstrates superior computational and storage efficiency compared to these methods. Specifically, even when compared to lightweight methods, IoT-AMLHP achieves a 97.18\% reduction in FLOPs, a 92.00\% reduction in the number of parameters, and an 89.40\% reduction in model size. By significantly reducing FLOPs, IoT-AMLHP minimizes computational overhead, enabling deployment on devices with limited processing capabilities without compromising performance. Moreover, IoT-AMLHP's compact model size of 0.32 MB allows deployment on various IoT hardware, including routers and gateways, without exceeding storage limitations.

In summary, the experimental results demonstrate that IoT-AMLHP not only achieves outstanding traffic classification performance but also exhibits the lowest computational complexity, number of parameters, and model size among all evaluated methods, making it highly suitable for deployment in resource-constrained IoT environments.

\begin{table}[!h]
\renewcommand{\arraystretch}{1.1}
\centering
\caption{Computational cost, parameter quantity, and model size.}
\label{Table::Lightweight}
\resizebox{0.6\linewidth}{!}{
\begin{tabular}{lccc}
\toprule
\textbf{Method} & \textbf{FLOPs (M)} & \textbf{Params (M)} & \textbf{Model Size (MB)} \\
\toprule[1pt]
1DCNN & 990.90 & 5.83 & 66.7 \\
ATTLSTM & 823.09 & 1.31 & 14.9 \\
CNNLSTM & 57478.15 & 2.40 & 27.4 \\
IoT-ETEI & 3124.00 & 10.38 & 118 \\
TSCRNN & 12614.07 & 2.89 & 33.1 \\
APPNET & 850.98 & 0.69 & 12.48 \\
Deeppacket & 1945.88 & 3.51 & 40.1 \\
LSTM & 34794.40 & 0.26 & \underline{3.02} \\
SAM & 2231.53 & 0.73 & 9.38 \\  
MATEC & \underline{248.87} & 2.60 & 29.7 \\
MISCNN+ & 2546.42 & \underline{0.25} & 3.09 \\
\midrule[1pt]
\multirow{2}{*}{IoT-AMLHP} & \textbf{7.01} & \textbf{0.02} & \textbf{0.32} \\
 & \multicolumn{1}{c}{($\downarrow{97.18\%}$)} & \multicolumn{1}{c}{($\downarrow{92.00\%}$)} & \multicolumn{1}{c}{($\downarrow{89.40\%}$)} \\
\bottomrule
\end{tabular}
 }
\end{table}

\subsubsection{Deployability analysis}

\begin{figure}[htbp]
  \centering
    \begin{minipage}[b]{\linewidth}
    \centering
      \subfigure[Time consumption]{
        \includegraphics[width=0.45\linewidth]{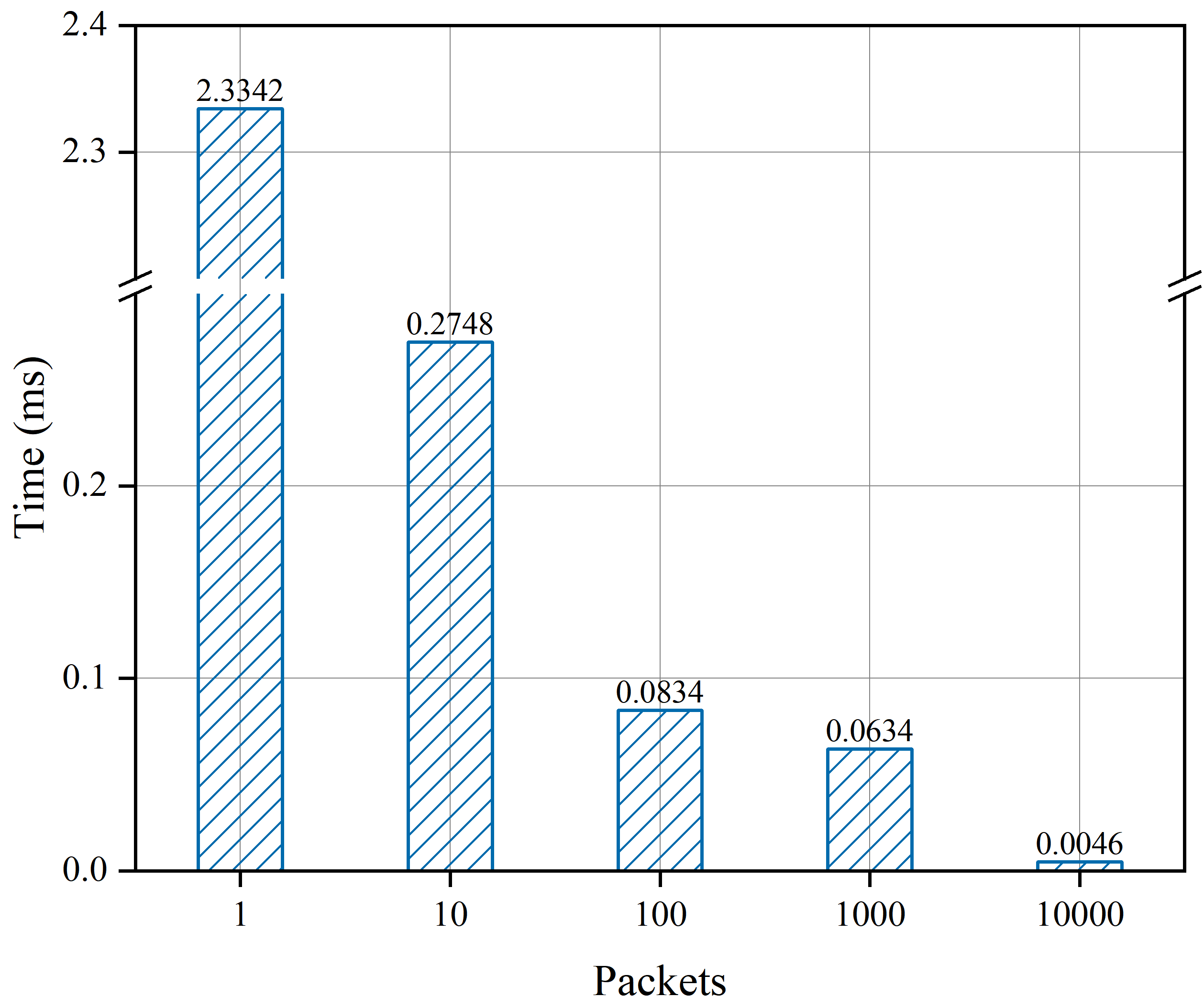}
      }
      \subfigure[Memory consumption]{
        \includegraphics[width=0.45\linewidth]{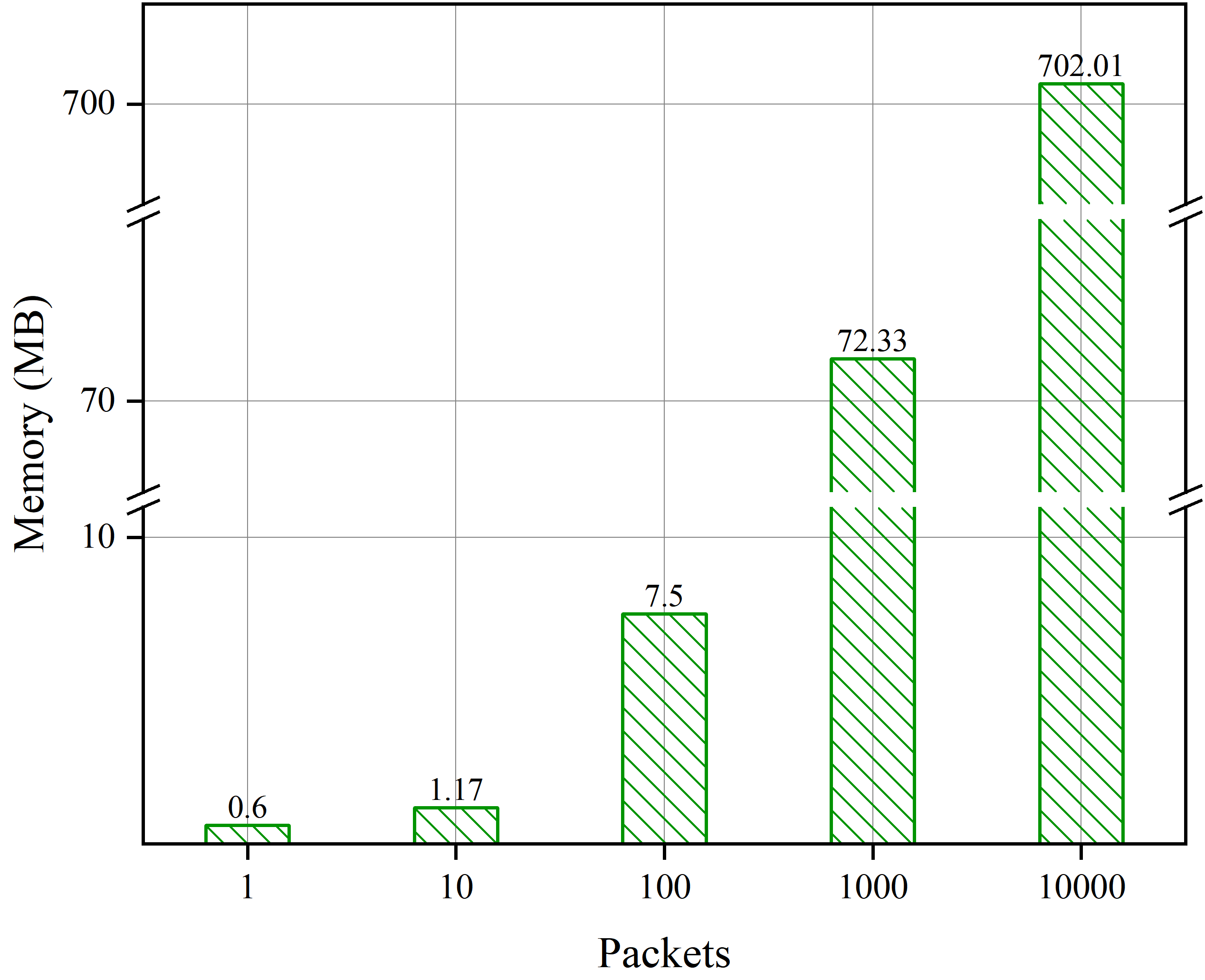}
      }
    \end{minipage}
    \caption{Inference time and memory consumption under different packets.}
    \label{fig::consumption}
\end{figure}

Although IoT-AMLHP exhibits exceptional efficiency in terms of computational and storage overhead, its practical deployment entails additional computational and memory demands due to the simultaneous inference of multiple packets and the intermediate tensors generated during this process. To validate the deployability of the proposed design, we evaluated IoT-AMLHP's runtime memory consumption and inference speed under different levels of parallel inference, as illustrated in Fig. \ref{fig::consumption}.

In terms of inference time, the average inference time per packet decreases substantially as the number of simultaneously processed packets increases. For instance, the average inference time for a single packet is 2.3342 ms, whereas processing 10,000 packets simultaneously reduces the average time per packet to 0.0046 ms. This trend demonstrates that parallel inference significantly enhances computational efficiency by leveraging the model's ability to process multiple packets concurrently. However, memory consumption increases notably with the number of packets processed in parallel. Specifically, memory usage grows from 0.6 MB for a single packet to 7.5 MB for 100 packets and 702 MB for 10,000 packets. This increase is due to the additional memory required to store intermediate features and tensors generated during parallel inference.

To strike a balance between inference time and memory consumption, we recommend processing 100 packets in parallel as the optimal configuration for IoT-AMLHP. At this level, the average inference time per packet is reduced to 0.0834 ms, while memory consumption remains at a manageable 7.5 MB. This configuration ensures efficient processing and resource utilization, making it particularly suitable for deployment on resource-constrained IoT devices such as edge gateways and routers.

The experimental results highlight IoT-AMLHP's adaptability to varying resource capacities. For devices with limited memory resources, such as low-power IoT devices, fewer packets can be processed in parallel to reduce memory usage. Conversely, for devices with more robust hardware, higher levels of parallel inference can be utilized to achieve faster processing speeds and higher throughput. This flexibility allows IoT-AMLHP to cater to diverse deployment scenarios, ensuring its practical utility in real-world IoT environments.

\section{Conclusions}  \label{sec::conclusions}

In this study, we propose IoT-AMLHP, an aligned multimodal learning framework for resource-efficient malicious IoT traffic classification. IoT-AMLHP first parses packet header and payload bytes to construct an aligned and standardized multimodal packet-wise representation. Then, the representation is fed into a resource-efficient neural network comprising a multimodal feature extraction module and a multimodal fusion module. The multimodal feature extraction module employs depthwise separable convolutions to extract multi-scale features from different modalities while maintaining a lightweight architecture. The multimodal fusion module adaptively focuses on complementary features from different modalities to effectively integrate the extracted multimodal representations. Through compact packet representations and lightweight neural network design, IoT-AMLHP ensures resource efficiency while achieving high accuracy for malicious IoT traffic.

The proposed method's efficacy was evaluated through extensive experiments on three public IoT traffic datasets. The results demonstrate the outstanding performance of IoT-AMLHP, achieving over 99\% malicious IoT traffic classification accuracy with only 7.01M FLOPs and 0.02M model parameters. Compared to other baselines, IoT-AMLHP exhibits superior malicious traffic classification capabilities while significantly reducing computational and spatial resource requirements. Furthermore, experimental results regarding inference time and memory consumption further validate the efficiency and practicality of IoT-AMLHP. Under reasonable resource allocation, IoT-AMLHP achieves an average classification time of approximately 0.08 ms per packet and a memory consumption of around 7.5 MB, indicating its ability to run efficiently on resource-constrained devices and adapt to various IoT scenarios. In future research, we will explore model compression techniques such as quantization and pruning to further reduce computational complexity and enhance the proposed method.

\section*{Declaration of competing interest}
The authors declare that they have no known competing financial interests or personal relationships that could have appeared to influence the work reported in this paper.

\section*{Data availability}
Data will be made available on request.

\section*{Declaration of generative AI and AI-assisted technologies in the writing process}
During the preparation of this work, the authors used GPT tools in order to improve the language. After using these tools, the authors reviewed and edited the content as needed, and take full responsibility for the content of the publication.

\section*{Acknowledgements}
This work was supported by the National Natural Science Foundation of China (Grants No. U2436601, U21B2003, 62072250, 61602247).

\printcredits

% \bibliographystyle{cas-model2-names}

% % Loading bibliography database
% \bibliography{}

% % Biography
% % Here goes the biography details.

% % Here goes the biography details.
\bibliographystyle{elsarticle-num} 
\bibliography{ref-AD}

\end{document}